\documentclass[twocolumn,prb,showpacs,preprintnumbers,amsmath,amssymb]{revtex4}
\usepackage{graphicx}
\usepackage{dcolumn}
\input epsf
\newcommand{\eqperiod}{\; \mbox{.}}
\newcommand{\eqcomma}{\; \mbox{,}}

\begin{document}
\title{Effect of the long-range interaction in transport through 
one-dimensional nanoparticle arrays}
\author{E. Bascones$^{1,2,3,*}$, J.A. Trinidad$^2$, V. Est\'evez$^1$ and 
A.H. MacDonald$^2$}
\affiliation{
$^1$Instituto de Ciencia de Materiales de Madrid, CSIC,Cantoblanco, E-28049 
Madrid, Spain 
\\
$^2$ Department of Physics, University of Texas at Austin, Austin TX-78712, 
USA
\\ 
$^3$Theoretische Physik, ETH-H\"onggerberg, CH-8050 Zurich, 
Switzerland}
\date{\today}
\begin{abstract}
We analyze the effect of the long-range interaction on the transport properties
through ordered and disordered one-dimensional metallic nanoparticle 
arrays. We discuss how the
threshold voltage, the I-V curves and the voltage drop 
through the array are modified as compared to the case in which interactions 
are restricted to charges placed on the same island. We show that some of these
modifications are due to finite interactions between charges in different 
nanoparticles while other ones are due to interactions between charges in the 
islands and those at the electrodes, what produces a polarization potential 
drop through the array. We study the screening of the disorder potential
due to charged impurities trapped in the substrate and find that long-range
interactions introduce correlations between the disorder potentials of 
neighboring islands.
\end{abstract}
\email{leni@icmm.csic.es}
\pacs{73.23.-b,73.63.-b,73.23.Hk}
\maketitle

\section{INTRODUCTION}

Nanoparticle arrays\cite{metallicwhetten,metallicheath,metallicjaegerdis,metallickiehl,metalliclin,metallicjaegerquasi1d,metalliczhang,metallicjaegernature,metallicschoenen,metallicysemi,semimurray,semitalapin,semibawendi,semisionnest,semidrndic,semiheath,magneticsun,magneticblack,magneticpuntes,mixedredl,mixedkorgel,mixedshevchenko} are a perfect system to analyze correlated electronic 
transport and have received a lot of  experimental and theoretical 
attention during the last decade\cite{singlecharge,collectiveheath1,collectiveheath2,collectiveheathpb,collectivebard,collectivekorgel,collectivepileni,collectivesheng,Ancona01,Tinkham99,Bakhalov89,Middleton93,Matsuoka98,Shin98,Stopa01,Kaplan02,Kaplan03,Kinkhabwala04,Schoeller05,DasSarma,Belobodorov,Glazman,Turlakov,Efetov,Shklovskii}.
In spite of  this, their properties and I-V characteristics 
are not well understood. 
The presence of disorder in these arrays complicates the 
analysis\cite{disorderjaeger,disorderheath,disordercordan} .
The two most relevant 
and experimentally unavoidable types of disorder are
charging disorder due to random
charges quenched at the substrate and resistance disorder due to small changes
in the distance between the nanoparticles and the exponential dependence of the
resistance on the distance. Voids in the lattice or a 
non-homogeneous distribution of nanoparticle size
can be also present\cite{metallicjaegerdis} but this source of disorder is 
less important in some 
recent
experiments.    

Theoretical analysis have mainly considered arrays in which each 
nanoparticle is capacitively coupled only to its nearest 
neighbors\cite{Bakhalov89,Middleton93,Hu94,Melsen97,Berven01,Kaplan03,Jha05}, 
especially the case in which this coupling is small. 
The truncation of capacitive coupling to nearest neighbor results in an 
interaction between charges in  
different conductors which decays exponentially with the distance between 
them\cite{Bakhalov89,Middleton93}.  This limit
is relevant for those arrays coupled to a gate electrode\cite{Clarke98}, 
as the 
mobile charges in this lead effectively screen Coulomb interactions.
A complete description of the non-equilibrium transport through arrays has 
been presented
only very recently\cite{ourpaper} and is restricted to one-dimensional 
metallic nanoparticle
arrays 
in which interactions are
finite only when charges are placed on the same conductor.

Self-assembled arrays fabricated nowadays are deposited onto insulating 
substrates and generally lack a gate voltage. In these arrays, the screening 
of 
long-range interactions is less effective, but the proximity of other 
conductors, both 
islands and leads, modifies its value compared to a $1/r$ Coulomb 
law\cite{Matsuoka95,Whan96}. 
The electrodes contribute to the screening of the interaction. 
Theoretical analysis including the effect of long-range interactions are
scarce and limited\cite{Kaplan03,Reichhardt03,Schoeller05}
to numerical results or particular cases. 
In this paper we perform a detailed analysis of  the effect of the long-range 
character of the
interaction  on the transport properties through ordered
and disordered one-dimensional arrays. 
We consider the influence of charging and resistance disorder.  
The interactions are described by an inverse
capacitance matrix, in which coupling to other conductors 
is not truncated. The matrix is 
calculated  including the effect of screening and the current is computed
numerically. Some analytical approximations are also discussed and compared 
with the numerical results. This comparison allows us to understand the origin 
of the behavior found. 
To analyze the transport we put forward a model to describe the 
electrostatic interactions among the charges occupying the islands and the 
electrodes, which allows for a  clear description of the relevant
contributions to the transport. In particular, we introduce the concept of
polarization potential drop at the junctions between the nanoparticles. 
As in our previous study of the short-range case\cite{ourpaper}, we analyze
the threshold voltage, flow of current and voltage drop for the cases with and
without charge or junction resistance disorder.

The organization of the paper is as follows. Section II describes the system 
under study and the model used to analyze the transport properties and compares
it with previous models used in the literature. In section 
III we discuss the correlations  introduced by the 
long-range character of the interaction in the disorder potential. 
Section IV, V and VI respectively 
analyze the threshold voltage, I-V characteristics and voltage drop through the
array. In section VII we summarize our results. Finally, in Appendix I we 
describe the
two methods employed to calculate the screened interaction between charges, 
and in Appendix II
we discuss the effect of screening on the distance-dependence of this 
interaction.

\section{THE MODEL}

We consider an array  composed of  
$N$ metallic spheres of radius 
$r^{isl}$ and center to center distance $2r^{isl}+d$. 
Throughout lengths are measured in units of $r^{isl}$ and 
energies
in units of the charging energy of an isolated nanoparticle 
$E_{c}^{isl}=1/(2C^{isl})$, with $C^{isl}$ the nanoparticle capacitance when
it is isolated. Here and in the following the electronic charge $e=1$.
As $d/r^{isl}$ decreases, the charges in an island feel 
more the
effects of charges in other islands. 
In this sense we 
say that decreasing the spacing $d/r^{isl}$ increases the range of 
interactions. Experimentally  arrays 
in which the metallic nanoparticles are capped with thiols have 
$d/r^{isl} \sim 0.5$. Arrays self-assembled via other types of molecules, 
like DNA, allow for the investigation of larger $d/r^{isl}$ values.

We model each nanoparticle with a continuum level
spectrum. The tunneling barriers which separate them have a resistance much
larger than the quantum of resistance and we treat the transport at the
sequential tunneling level\cite{singlecharge}. 
To analyze the transport the array is sandwiched between two large electrodes.
We assume that the electrodes are not ideal voltage
sources, but have finite self-capacitances. The potential on the leads will 
thus fluctuate in response to all tunneling events, even those that do not 
directly involve the electrodes. For a more extensive discussion of these 
assumptions see\cite{ourpaper}

The energy of the system is
given by
\begin{equation}
F=\frac{1}{2}\sum_{\alpha,\beta=0}^{N+1} Q_{\alpha}C^{-1}_{\alpha
  \beta}Q_{\beta}+\sum_{i=1}^N Q_i \phi_i^{dis}
\eqperiod
\label{freecharge}
\end{equation}
Labels $0$ and $N+1$ refer to source and drain electrodes and $1,\ldots, N$ to 
the 
islands.
Latin capital and lower case letters
 are used to denote electrodes and islands respectively. Greek indexes will be 
used when the labels refer to both islands and electrodes.
$\phi_i^{dis}$ is a random potential at each island, present in charge 
disordered arrays, and zero in the clean case. Charges
$Q_0$ and $Q_{N+1}$ maintain source and drain electrodes at potentials 
$V_0$ and $V_{N+1}$, respectively.

The electrostatic interactions in our system are defined through an 
inverse capacitance matrix $C^{-1}$ that directly includes  
all the conductors in our system. 
All the elements $C^{-1}_{\alpha,\beta}$ are positive. 
The inverse capacitance matrix is symmetric, 
$C^{-1}_{\alpha \beta}=C^{-1}_{\beta\alpha}$, and has dimension 
$(N+2) \times (N+2)$, i.e. it includes the electrodes. 
We have developed two numerical methods 
to calculate the interaction potential (inverse capacitance matrix) 
of an array of spheres. These methods 
are explained in Appendix I.
The properties of this interaction, how it compares to previous calculations
for the case of cubic, cylindrical (with array and island axis collinear) and
disk shaped islands (with array and island axis perpendicular) and the
screening effect of the electrodes 
are discussed in
Appendix II. For simplicity we have considered two large spherical leads.
The coupling between the islands and the electrodes is included 
through $C^{-1}_{i,K}$ and $C^{-1}_{K,i}$.

The current is computed numerically by means of a Monte Carlo simulation,
described in\cite{ourpaper,Bakhalov89}. It is controlled by the probability of a tunneling
process, given by
\begin{equation}
\Gamma(\Delta E)=\frac{1}{R}\frac{\Delta E}{exp(\Delta E/K_B T)-1}
\end{equation}
Here $R$ is the resistance of the junction through which tunneling takes 
place and $\Delta E$ is the difference between the energy of the system before
and after the tunneling event. In the following we restrict the discussion to
zero temperature. Then $\Gamma(\Delta E)=-\Delta E/R_T \Theta (-\Delta E)$. It is finite only when $\Delta E$ is negative.

After a bit of algebra, the change in energy
due to a tunneling event from $\alpha$ to $\beta$ can be rewritten as
\begin{equation}
\Delta E = E^{e-h}_{\alpha,\beta}
+ (\phi_{\beta} - \phi_{\alpha})
\label{change} \eqperiod
\end{equation}
Here, the excitonic energy $E^{e-h}_{\alpha,\beta}$ necessary to 
create an electron-hole pair in an uncharged array is given by
\begin{equation}
E^{e-h}_{\alpha,\beta}=\frac{1}{2}C^{-1}_{\alpha \alpha}+
\frac{1}{2}C^{-1}_{\beta\beta}
-C^{-1}_{\alpha \beta} 
\label{excitonic} \eqperiod
\end{equation}
The second term in (\ref{change}) can be seen as the potential difference 
between the sites involved in the tunneling. At the electrodes 
$\phi_0=V_0=\alpha V$ and $\phi_{N+1}=V_{N+1}=(\alpha-1)V$, where $V$ is the
bias potential and we have introduced the bias asymmetry parameter 
$\alpha$ as in ref.\cite{ourpaper}. 
Both $\alpha=1/2$, also denoted as symmetric bias, and $\alpha=1$ have been 
used in the literature. 
$\alpha$ characterizes how the bias 
voltage is partitioned between source and drain chemical potential shifts.
Since no physical properties depend on the 
overall 
zero of energy, varying $\alpha$ in our model is entirely equivalent to 
rigidly shifting all impurity potentials by $- \alpha V$.
Since in our model all transport occurs by 
transfer between adjacent nanoparticles, the evolution of a nanoparticle array 
as the bias voltage is applied is sensitive to $\alpha$, and we believe that 
the dependence on $\alpha$ could in principle be observable, see discussion in
\cite{ourpaper}.

At the islands the potential can be decomposed 
into three terms $\phi_i=\phi_i^{dis}+\phi_i^{pol}+\phi_i^{ch}$, the disorder
potential $\phi_i^{dis}$ due to random charges in the substrate , 
the polarization potential  $\phi_i^{pol}$ at
the island induced by the electrodes at finite bias and the potential
due to the charges in the nanoparticles $\phi_i^{ch}$. 
Here 
\begin{equation}
\phi^{pol}_i=\lambda^{\alpha}_i V
\label{phipol}
\end{equation} 
with
\begin{eqnarray}
\nonumber
\lambda^{\alpha}_i=C^2_{gen}\left[\alpha \left( C^{-1}_{i0}C^{-1}_{N+1,N+1}-
C^{-1}_{i,N+1}C^{-1}_{N+1,0}\right ) 
\right.
\\
\left.
+ (\alpha -1)\left( C^{-1}_{iN+1}C^{-1}_{00}-C^{-1}_{i0}C^{-1}_{N+1,0}\right )
\right ] \eqperiod
\label{lambda2}
\end{eqnarray}
and 
\begin{equation}
C^{2}_{gen}=\frac{1}{C^{-1}_{00}C^{-1}_{N+1,N+1} -(C^{-1}_{N+1,0})^2}
\eqperiod
\label{cgen}
\end{equation}
The charging potential
\begin{equation}
\phi^{ch}_{i}=\sum_{j=1}^N Q_j \tilde C^{-1}_{ij}
\label{chargingpot}	
\end{equation}
with 
\begin{eqnarray}  
\label{ctilde}
\nonumber
\tilde C^{-1}_{ij}=C^{-1}_{ij} + 
C^2_{gen}\left [C^{-1}_{0,N+1}\left 
(C^{-1}_{i N+1}
  C^{-1}_{j0} + C^{-1}_{i0}C^{-1}_{j,N+1}\right )
\right .
\\ 
\left .
 - C^{-1}_{00}C^{-1}_{N+1,i}C^{-1}_{j,N+1}-C^{-1}_{N+1,N+1}C^{-1}_{i
  0}C^{-1}_{j0}\right]
\eqperiod
\end{eqnarray}
$\tilde C^{-1}$ can be interpreted as a modification of  the 
interaction between the charges in the islands due to the proximity 
of the electrodes 
at a fixed potential. For the case $i=j$ in which both charges are on the 
same
island this modification was already discussed in\cite{Bakhalov89}, as the 
interaction of a soliton with a passive edge. Expression (\ref{ctilde}) 
shows that not only when the charges are in the same island, but also when 
they occupy different islands, their effective interactions are modified by the
presence of the voltage-biased leads.
Two types of terms 
can
be differentiated in the modification of this interaction. The last two terms 
in (\ref{ctilde}) or direct terms, can be viewed as the interaction 
between a charge in island $i$
and the image charge at one of the electrodes induced by the charge in island 
$j$. This term is
 affected by the presence of the other electrode. 
On the other hand, the terms containing $C^{-1}_{0 N+1}$, or indirect terms, 
reflect the interaction between the image charges in both electrodes.   
Direct and indirect terms have opposite sign. The direct term reduces 
the effective interaction; the indirect one increases it.
We emphasize that (\ref{change}) to (\ref{ctilde}) follow from 
(\ref{freecharge}) after straightforward and trivial algebra. We have just 
defined a few quantities and split the change in energy $\Delta E$ and 
potential $\phi_\alpha$ in several terms to facilitate the physical 
interpretation of the transport properties.

We can also define the potential drop at each junction
\begin{equation}
\Phi_i=\phi_{i}-\phi_{i-1}
\label{phijunction}
\end{equation}
with the corresponding disorder, polarization and charging terms $\Phi^{dis}$, 
$\Phi^{pol}$, and $\Phi^{ch}$. 
Label $i$ for a junction runs from $1$ to $N+1$ and  refers to the one between 
conductors $i-1$ and $i$.
The polarization potential drop at each junction,
\begin{equation}
\Phi^{pol}_i=\Lambda^\alpha_iV=(\lambda_i^\alpha-\lambda_{i-1}^\alpha)V \eqcomma
\label{lambdajunction}
\end{equation} 
does not depend on the resistance of the junctions, but on the electrostatic 
interactions of each island with the voltage-biased
leads. Here $\lambda_0^{\alpha}=\alpha$ and $\lambda_{N+1}^\alpha=\alpha-1$. 

A linear drop of the polarization potential implies
$\phi^{pol}_i=\left (\alpha -\frac{i}{N+1}\right ) V$, requires 
$\Lambda^\alpha_i=1/(N+1)$ for all junctions 
and independence on $\alpha$. 
From (\ref{lambda2}) 
we see that a priori $\lambda_i^{\alpha}$ 
depends both on the geometry of the electrodes and the array 
and on how this is biased. It depends on how 
the charges in the
islands and the electrodes interact.
For most capacitance matrices, in particular for the 
capacitance matrices discussed here, the polarization potential is not 
linear in the island label $i$ 
and the potential drops are larger close to the biasing leads.
In the onsite case, discussed in ref.\cite{ourpaper} 
$\Lambda^{\alpha\{onsite\}}_i$ 
is finite only at junctions $1$ and $N+1$ 
and 
given by $\Lambda^{\alpha\{onsite\}}_1=\alpha$ and 
$\Lambda^{\alpha\{onsite\}}_{N+1}=\alpha - 1$. 
In general, as the range of the interactions between charges 
increases,  
$\Lambda^{\alpha}_{i}$ is more homogeneous

In previous models\cite{Middleton93} $C^{-1}$ had dimension $N \times N$ and 
the inverse self-capacitances
 of the electrodes, $C^{-1}_{00}$ and $C^{-1}_{N+1, N+1}$, and the 
 inverse mutual capacitances between them, $C^{-1}_{0,N+1}=C^{-1}_{N+1,0}$, 
were neglected. 
In 
our model $C^{-1}_{K,L}$ are small quantities whose values will not
significatively affect the quantitative results.
These other models do not include the indirect term in (\ref{ctilde}) either. 
On the other hand, an expression analogous to (\ref{phipol}) can be defined in 
other models. For
example, in
 the model previously discussed by Middleton and Wingreen\cite{Middleton93}, 
with an $N \times N$ capacitance matrix, the interaction between an island 
and an 
electrode is given by the 
interaction of the charge of the island with charges induced by the electrode 
in the islands immediately adjacent to the electrode. This interaction
results in a polarization potential characterized by
\begin{equation}
\lambda^{MW,\alpha}_i=C_{i-el}\left ( \alpha C^{-1}_{1i}-(\alpha -1)C^{-1}_{Ni}
\right )
\label{lambdaMW}
\end{equation}            
Here $C_{i-el}$ is the capacitance between the source or drain electrode and 
an adjacent island. In this model, except in the extreme long-range case, 
see below, the polarization potential does not decay linearly with distance 
and 
depends on the asymmetry of the bias potential. 
Other models, however, impose a uniform polarization drop through the array
\cite{Reichhardt03} 

As discussed in\cite{ourpaper} and section VI, even if the polarization 
potential does not
drop linearly and is independent on the junction resistance, the average total
potential drop depends on the resistance (via the average charge occupation of
the islands) and for homogeneous resistances a linear drop is partially 
recovered at large bias voltages.

Whenever not specified we assume that all the junction resistances $R_i$ 
are equal 
and given by $R_T$. The effect of non homogeneous resistances will be studied 
in two ways. In the first case, one of the junction resistances at a given 
position is larger than
 the other ones (given by $R_T$). In the second case the value of the 
resistances, varying in between two values 
is randomly assigned to the junctions. To mimic
that disorder in resistances originates in variations in distances between the
islands and the exponential dependence of the junction resistance in the 
distance between islands the 
junction resistance is given by  $R=R_0 exp(\gamma dist)$ with $R_0$ and 
$\gamma$ input parameters and $dist=1 + random/2$. Here $random$ is a random 
number between $0$ and $1$. In the paper, we have used $R_0=1.1825 R_T$ and 
$\gamma=1.526,1.95,2.84$. With these values the resistance changes 
respectively 
between (5-11)$R_T$, (8-21)$R_T$ and (23-83)$R_T$.

\section{Screening of Disorder Potential}
\label{disorder}
Charged impurities trapped in the substrate underlying the nanoparticle array 
create random potentials at the nanoparticles. In molecularly assembled arrays,
 charge transfer to the organic molecules surrounding the nanoparticles 
results in non-integer random charges 
at the islands\cite{Xue03}. 
Charging disorder is included in our model through a random potential at 
each 
island $\phi^{dis}_i$.
In principle, $\phi^{dis}_i$ 
take values larger than the charging energy $E_c^{isl}$. However, for large 
values 
of the disorder potential, charges flow to compensate for these large 
fluctuations. 
In this section we analyze the effect of the long-range 
interaction on the final distribution of disorder potential as compared
to the case with onsite interactions. We find that 
the distribution of probabilities $P(\phi^{dis})$ and $P(\Phi^{dis})$ are 
modified.
The maximum and minimum values of $\{\phi_i^{dis}\}$ and $\{\Phi_i^{dis}\}$
are modified compared to the short range case and given by 
$\pm C^{-1}_{ii}/2$ and $E^{e-h}_i$. Correlations between the disorder 
potentials of neighboring islands are introduced.

If interactions between the charges are short-range, 
($C^{-1}_{ij}=\delta_{ij}$), the set of disorder potentials 
$\{\phi^{dis}_i\}$, 
once the screening of the potential due to the mobile charges is taken into 
account, is uniformly distributed  in the interval 
$-E_c^{isl} \leq \phi_i^{dis} \leq E_c^{isl}$.  
The probability associated with each pair, 
$(\phi^{dis}_{i},\phi^{dis}_{i-1})$ ,
is a constant, see Fig.~1 and the distribution of the 
probabilities of the potential drops due to disorder across the array 
junctions, $\Phi^{dis}_i = \phi^{dis}_i -\phi^{dis}_{i-1}$, has the 
form\cite{disorderjaeger}
\begin{equation}
P(\Phi^{dis})=\frac{1}{\Phi^{dis}_{MAX}} \left( 
1-\frac{|\Phi^{dis} |}{\Phi^{dis}_{MAX}} \right) 
\label{pdmuonsite} 
\end{equation}
and $\Phi^{dis}_{MAX}=2 E_c^{isl}$.
In the presence of long-range interactions, the charges 
which flow to compensate the large fluctuations of the disorder potential, 
influence the value of the total potential at neighboring islands. As a 
consequence, the screened disorder is correlated\cite{disorderjaeger}. 
The probability of 
each pair $(\phi^{dis}_{i},\phi^{dis}_{i-1})$ is no longer  a constant. 
$P(\Phi^{dis})$  depends on the inverse capacitance matrix $C^{-1}$. 
In order to analyze these correlations and obtain the proper disorder 
potential distribution we assign the potentials by first randomly assigning 
potentials 
to the islands $\phi_i^{dis-bare}$, in the interval 
$-W \leq \phi^{dis-bare}_i \leq W$ with $W$ larger than the charging energy. 
We then find the equilibrium configuration of charges $\{Q^{sc}_j\}$ that 
occupy the array with island disorder potentials $\{\phi_i^{dis-bare}\}$ and 
grounded leads ($V_0 = V_{N+1} = 0$) and redefine the 
potentials at each site using the expression
\begin{equation}
\phi^{dis}_i= \sum_{j=1}^{N} \tilde C^{-1}_{ij} Q^{sc}_j + 
\phi^{dis-bare}_{i} \eqperiod
\label{screeneddis}
\end{equation}
The effect of the screening charges $\{Q^{sc}_j\}$ is included in the 
redefined potentials $\{\phi^{dis}_i\}$ so we then reset the number of 
charges at each site to zero to avoid doublecounting the charge when we 
calculate the total electrostatic energy of our system. 

Following the redefinition of the disorder potentials, we find that
on average the distribution of the 
disorder potentials $\{\phi^{dis}_i\}$ and the disorder potential drops 
$\{\Phi^{dis}_i\}$ between adjacent islands are independent of $W$.
The values of  $\{\phi^{dis}_i\}$ and  $\{\Phi^{dis}_i\}$  are bound by
$\pm C^{-1}_{ii}/2$ and $\pm E^{e-h}_{i}$ respectively, as 
the total energy of the system
is at a global minimum when the original disorder configuration
$\{\phi_i^{dis-bare}\}$ is screened out by the charges $\{Q^{sc}_j\}$ .
When in this state, adding an additional charge to any island
in the array increases the energy of the system.  The 
energy of adding an additional charge to an island from a large
electrode outside the system with neglible self inverse capacitance
is given by $E_i^{add}= (1/2)C^{-1}_{ii} \pm \phi^{dis}_i$ where
the top (bottom) sign refers to the change in energy of the system
as a result of adding a positive (negative) charge to island $i$.  
Since $E_i^{add} > 0$ when the array is in equilibrium, the
disorder potential values must lie between $\pm  (1/2)C^{-1}_{ii}$.
The magnitude of this  quantity equals $E_c^{isl}$ in the  onsite limit and
decreases as the strength and range of the Coulomb interactions
increases, see Appendix II.
Additionally when the array is in equilibrium state, the energy to
hop between all pairs of adjacent sites must be greater than zero.
From (\ref{change}), the disorder potential differences $\Phi_i^{dis}$ are
restricted between $\pm E^{e-h}_i$.  $E^{e-h}_i$ equals $2E_c^{isl}$
in the onsite limit and decreases with decreasing $d/r^{isl}$, as shown in
Fig. 8.

\begin{figure}
\includegraphics[width = 3.3 in ]{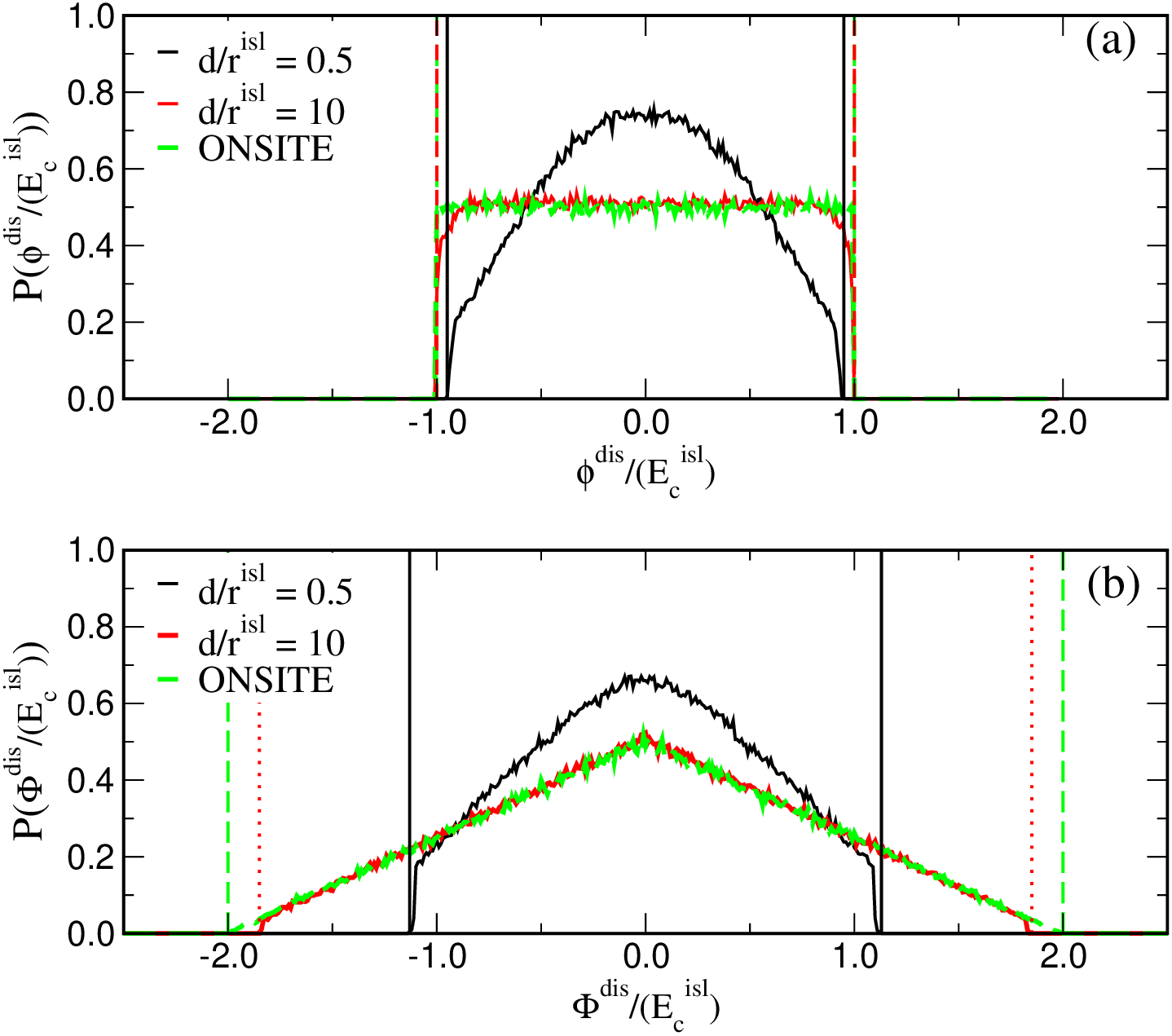}
\caption{Probability distributions of disorder potentials $\phi^{dis}$ in (a),
and disorder potential differences $\Phi^{dis} $ in (b), 
due to disorder for 50-island arrays with purely onsite
interactions and with long range interactions at two spacings: 
 $d/r^{isl} = 0.5$ and 10.  
The onsite $\Phi^{dis}$ distribution
is described by Eq.(\ref{pdmuonsite})
with $\Delta \phi^{dis}_{MAX} = E^{e-h, onsite}_{i,i-1} = 2E_c^{isl}$
where $E_c^{isl} = 1/(2C^{isl})$.   For all the cases,
$P(\phi^{dis})$ ($P(\Phi^{dis})$)
is finite valued for $\phi^{dis}$ ($\Phi^{dis}$) between 
$\pm 0.5 C^{-1}_{ii}$
($\pm E^{e-h}_i$) which decreases with decreasing spacing. 
As the spacing decreases, the probability of having 
$\Phi^{dis}$ and $\phi^{dis}$ values close to zero increases.
Vertical lines are included as guidelines to emphasize the edges of
the distributions.
}
\label{dmufig}
\end{figure}

Fig.~1(a) and (b) compare $P(\phi_i^{dis})$ and $P(\Phi_i^{dis})$ 
for arrays with
purely onsite interactions ($C^{-1}_{i \ne j} = 0$)  with arrays with long 
range interactions 
($C^{-1}_{i \ne j} \ne 0$) at two spacings, $d/r^{isl} = 0.5$ and 
$d/r^{isl} = 10$.  
In all cases, $\phi_i^{dis}$ and $\Phi_{i}^{dis}$ are calculated for arrays 
with 
50 islands in between two grounded leads.  The histograms average the values
of the potentials of all islands and the values of the potential drops between 
all adjacent
islands over many realizations of disorder $(O(>10^4))$.  The smaller spacing, 
$d/r^{isl} = 0.5$,
is typical of chemically assembled nanoparticle arrays.  
The larger spacing, $d/r^{isl} = 10$, is
atypical of arrays in most experiments but is mainly 
included as a pedagogical example 
because it
has interactions among islands that are finite yet comparable to the onsite 
case
that is often used to describe experiments \cite{Middleton93,metallicjaegerdis,
disorderjaeger}.   
Arrays recently synthesized\cite{metalliczhang} have large $d/r^{isl}$ values.
In Fig.~1(a), $P(\phi_i^{dis})$ is
a constant between $\pm E_c^{isl}$ for the onsite case.  As the range of
interactions increases (decreasing $d/r^{isl})$, the width of $P(\phi_i^{dis})$
decreases because the disorder potential values are
bound by $\pm 0.5 C^{-1}_{ii}$.  Increasing Coulomb interactions
also increases (decreases) the probability of small (large) values
of $|\phi^{dis}_{i}|$ .  In Fig.~1(b), the onsite $P(\Phi^{dis}_i)$
distribution is given by (\ref{pdmuonsite}).  Similar to the trends in 
Fig.~1(a),
as the range of Coulomb interactions increases, the width of
the distribution decreases and the probability of small (large) 
$|\Phi^{dis}_i|$ values
increases (decreases).  The increased probabilities of small 
$|\Phi^{dis}_i|$ are
due to Coulomb correlations that make it more likely for the disorder 
potentials
of neighboring islands to have similar values.  See Fig.~2.
Increasing the range of Coulomb interactions
leads to a greater relative reduction in  the width of $P(\Phi^{dis}_i)$ than 
$P(\phi^{dis}_i)$ because the former are bound by $\pm E^{e-h}_i$
whereas the latter are bound by $\pm 0.5 C^{-1}_{ii}$. 
In the onsite case, $E^{e-h}_{i}$ equals $2E_c^{isl}$ for
all junctions between two islands and increasing Coulomb interactions
can reduce $E^{e-h}_i$ significantly due to an decrease (increase)
in $C^{-1}_{ii}$ ($C^{-1}_{i,i \pm 1}$).  See (\ref{excitonic}).

Our results for $P(\Phi_i^{dis})$ differ to some extent 
from those by Elteto et al\cite{disorderjaeger}, calculated with
an inverse capacitance matrix $C^{-1}_{ij}$ that is finite only for 
nearest neighbors and charge disorder modelled by a set of stationary quenched
charges. Elteto et al\cite{disorderjaeger} distributions are bound by 
$\pm C^{-1}_{ii}$ instead of by $E^{e-h}_i$ due to the lack of correlations
in their quenched disorder model. We permit the interactions among the 
screening charges to determine whether or not the disorder potentials are 
correlated.  

\begin{figure}
\includegraphics[width = 3.3 in ]{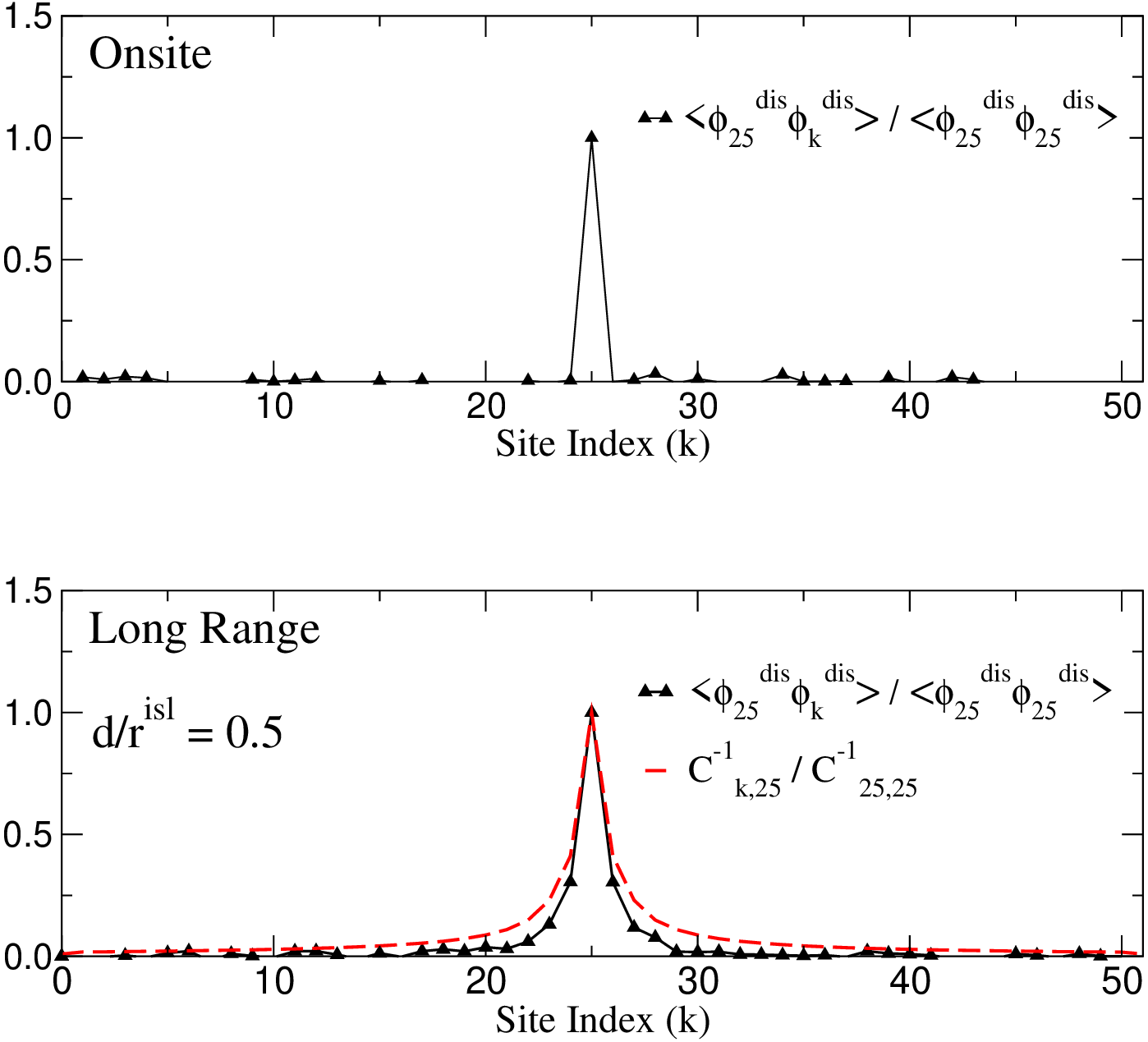}
\caption{
Comparison of $<\phi^{dis}_{25} \phi^{dis}_{k}>$ normalized by 
$|\phi^{dis}_{25}|^2$
for 50-island  arrays with onsite (top plot) versus long range (bottom plot) 
Coulomb interactions. 
In the presence of long-range interactions between the charges, the values of 
the disordered potentials are also correlated.
$C^{-1}_{k,25}$ normalized by $C^{-1}_{25,25}$ is 
included in
the long range case to show that correlations in the disorder potentials
are related to, but decay faster than the $C^{-1}$ elements.
}
\label{vivjfig} 
\end{figure}

In Fig.~2, we plot $<\phi^{dis}_i\phi^{dis}_k>$ to
show how interactions among charges affect the correlations
among disorder potentials.
In the onsite case, correlations are finite only if $i=j$.  In this case,
the disorder potentials of different islands are uncorrelated. 
In the case of 
long-range interactions with $d/r^{isl}=0.5$, correlations are maximal when
$i=j$, but they
do not vanish for $i \neq j$.  $<\phi^{dis}_i\phi^{dis}_k>$ is finite for at 
least $|i-k| \leq 3-4$. The correlation of disorder decays faster than the 
interactions as shown in the figure.  The correlations between $\phi^{dis}_i$
and its nearest neighbors $\phi^{dis}_{i \pm 1}$ make it more likely for
the disorder potential differences $\Phi^{dis}_i$ to have small magnitudes.

\section{THRESHOLD}
\label{threshold}
The threshold voltage is the minimum voltage which allows the flow of current.
In this section we analyze the threshold voltage of one-dimensional arrays with
long-range interactions for both clean and disordered systems. As in the 
short-range case the threshold is completely determined by energetic 
considerations and is independent of junction resistance disorder. In the clean
case we find that the threshold equals the minimum voltage necessary to create 
an electron-hole pair and increases with the number of particles at a rate 
which
depends on $d/r^{isl}$. For charge disordered arrays the average threshold is 
reduced compared to the onsite interactions case,
and increases  linearly with the number of islands, with a slope that 
decreases with decreasing $d/r^{isl}$, i.e. with increasing the range of the 
interaction.

When the interaction strength between charges at the nanoparticle and those at 
the electrodes does not vanish for any particle the  
polarization potential drop at every junction is finite.
In a clean array, the potential gradient created by this polarization 
potential drop allows current once an electron-hole pair is created, opposite 
to what happens in the onsite case. As a 
result,
the threshold voltage equals 
the mimimum voltage which allows the creation of an electron-hole pair.

The cost in energy to create an electron-hole pair in junction $i$ in an 
uncharged array (i.e. an array in which the nanoparticles have no excess 
charges) is 
$\Delta E=E^{e-h}_i -\Lambda_i^\alpha V$. We can define a junction dependent 
threshold voltage for creating an electron-hole pair 
$V^{TH,\alpha}_i=E^{e-h}_i/\Lambda_i^{\alpha}$.  In the onsite limit 
$V^{TH,\alpha}_i$ is finite only
at one or both contact junctions and infinite at the bulk, but with long-range 
interactions $V^{TH,
\alpha}_i$ is finite at every 
junction. For those cases analyzed, we have found that due to the 
smaller value of the
excitonic energy 
and the larger potential drop 
$V^{TH}_i$ is smallest at the contact junctions and
the threshold voltage is  
controlled by them.

Figs.~3(a) and (b) 
show the dependence of the threshold voltage $V_T$
of clean, symmetrically biased  arrays with long-range interactions on
the number of islands in the array $N$ and on the 
spacing between array sites,  $d/r^{isl}$.
The threshold voltage is determined by those factors that
define the polarization
potential drops across the contacts, $\Lambda^{\alpha}_1$ and 
$\Lambda^{\alpha}_{N+1}$.
$V_T$ increases with increasing $N$
because the fraction of the  polarization potential which drops 
across the contact junctions decreases as $N$ increases.  
As the spacing between the leads increases,
the polarization potential drop across each contact decreases  
until eventually it reaches a minimum value 
at which the polarization of each contact is only due to the interaction of 
each contact with the lead closest to it.  As a result, $V_T$
increases sublinearly with increasing $N$ and eventually
saturates.
For $N$ and $d/r^{isl}$ large enough that the polarization
potential drop across the contacts is not strongly influenced by interactions 
with the opposite lead, decreasing the array spacing decreases
the polarization potential drop across the contact junctions and
the threshold increases.
For $N$ and $d/r^{isl}$ small enough that both
leads strongly influence the polarization of both contact junctions,
decreasing the spacing increases the polarization potential drop
across the leads and the threshold decreases. 
The potential drops and threshold can be estimated by using an unscreened
$r^{-1}$ model for the inverse capacitance elements associated
with the leads, $C^{-1}_{i,0}$ and $C^{-1}_{i,N+1}$. 
These estimates
are included as dashed lines in Figs.~3(a) and (b).

\begin{figure}
\includegraphics[width=3.3 in]{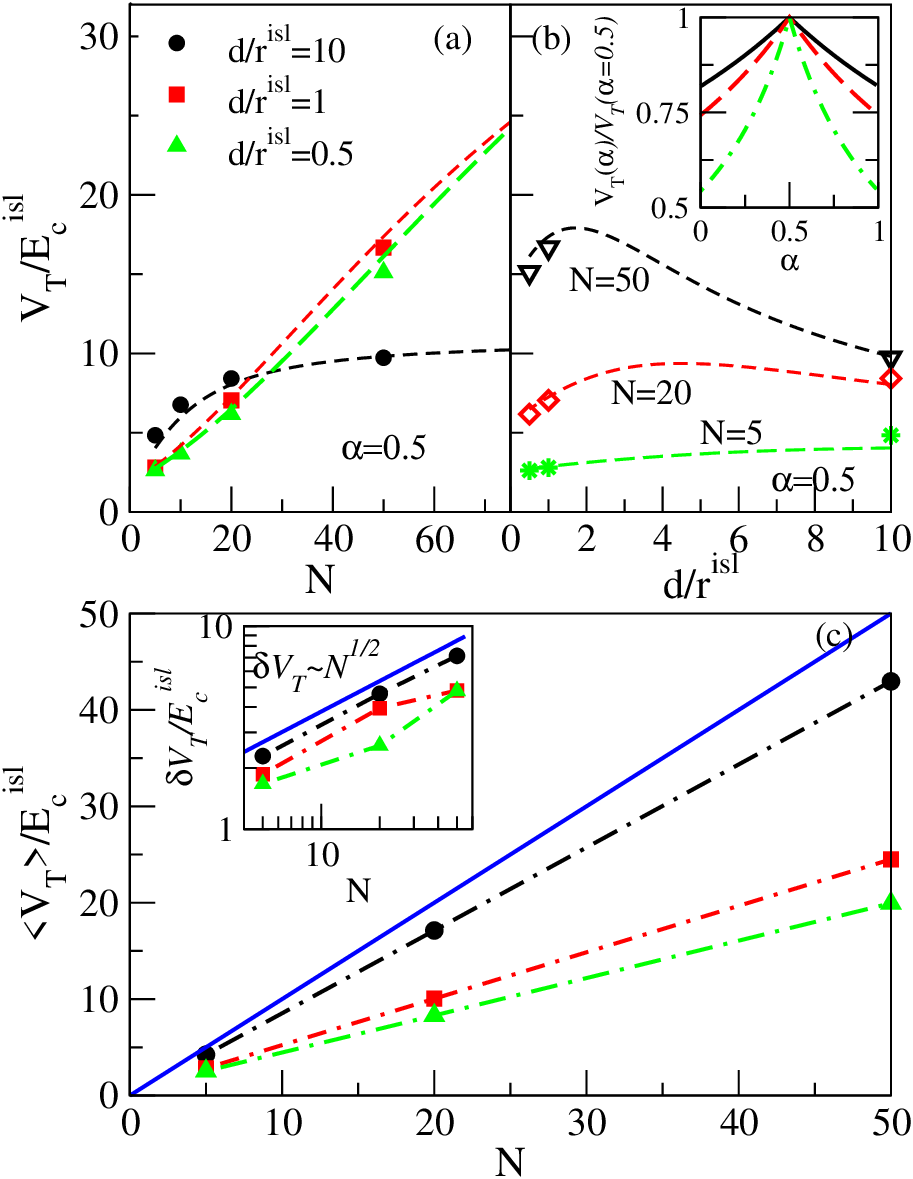}
\caption{(a) and main figure in (b): With symbols it is respectively plotted 
the threshold 
voltage $V_T$ of symmetrically biased arrays $\alpha=0.5$, with no disorder as 
a function 
of number of islands $N$ and of array spacing $d/r^{isl}$ for the inverse 
capacitances calculated as described in appendix I. As decribed in the text, 
the threshold voltage of clean arrays is controlled by the value 
$\Lambda^\alpha_i$ at 
the contact junctions. The dashed lines are
 estimates of the threshold voltage that use a $r^{-1}$ interaction to 
approximate the
 polarization potential drops across the contact junctions $\Lambda^\alpha_1$ 
and 
$\Lambda^\alpha_{N+1}$. Inset in (b): Threshold voltage of clean arrays for 
several 
array parameters as a function of $\alpha$ normalized to the value for a 
symmetrically biased array. From top to bottom $d/r^{isl}=0.5, N=50$; 
$d/r^{isl}=0.5, N=20$; and $d/r^{isl}=10, N=50$. (c) Main figure: Average 
threshold voltage of disordered arrays versus the array length at three 
different array spacings. The solid line shows the dependence of the average 
threshold on array length in the limit of onsite interactions. The inset shows 
the root mean square deviation of the threshold voltage distribution. For small 
$d/r^{isl}$ it deviates from the $N^{1/2}$ dependence found in the onsite 
case (in solid line). In inset and main figure dashed-dot lines are a guide to the eye and same legend as in (a) applies}  
\end{figure}

As shown in the inset of Fig.~3(b) 
$V_T$ changes smoothly with $\alpha$, contrary to the 
peak-valley structure found in the onsite interaction 
case\cite{ourpaper}. 
The threshold voltage depends  less on $\alpha$ as $N$ increases and as 
$d/r^{isl}$ decreases because  the applied voltage drops more homogenously 
across 
the array junctions.
This dependence disappears completely if the polarization potential 
drops linearly. In this last case the threshold voltage of clean arrays would 
be $V_T=(N+1)E_c^{isl}$. This threshold is linear in the number of particles, as in 
the onsite case, but its origin of  linearity and its slope differs in 
both cases.  

In the charge disordered onsite case every up-step in the disorder potential
prevents current flow and has to be compensated by a charge gradient.
As a result, the threshold voltage is controlled by  the number of 
up-steps, but this 
is not
the case for long-range interactions. 
In the long-range case the situation is more complex. Due to the finite
polarization potential drop at the inner junctions the number of junctions
which prevent the flow of current is reduced, compared to the onsite case. 
Up-steps in the disorder potential drop 
can localize the charge only if its value is larger than the polarization 
potential
 drop at the same junction. 
In some cases in which there is finite polarization drop at a 
junction, that is slightly smaller than the energy cost for tunneling it is 
possible that a small increase in the voltage in the electrodes permits the 
tunneling.
Increasing the voltage above the minimum bias voltage  
which permits the  generation of electron-hole pairs will, 
in some cases, but not always, 
result in a negative potential drop at the given junction allowing the flow 
of charges. But, quite often, the entrance of more charges and the creation of 
a charging potential gradient will be required in a similar way as it happened 
in the onsite case. Note that due to the polarization voltage drop at the bulk,
the threshold junction can be other than the contact ones.
Moreover, the interaction between charges
 in different islands decreases the energy for the entrance of a charges 
with opposite sign and increases the one for the entrance of charges with the 
same sign.  This effect was 
attributed to an attraction (repulsion) between the injected soliton and an 
antisoliton (soliton) on the array by 
 Likharev and coauthors\cite{Bakhalov89}.
Accumulation of charges in the array increases the threshold voltage. On the 
other hand a value of $\phi^{dis}_i$ at the contact islands favourable for the
entrance of charge unto the array can decrease it, as the polarization 
potential drop to allow entrance of charge is smaller. Both mechanisms 
compete to 
determine $V_T$. For large $d/r^{isl}$ the accumulation of 
charges is more important as 
the voltage drop at the bulk junctions is small and on average $V_T$ of large 
arrays will increase compared to the clean array threshold voltage. On the 
contrary the second effect can be more important for small $d/r^{isl}$.

Numerically we have found a linear dependence of the threshold
voltage on the number of particles in the array, see 
Fig.~3(c).
Decreasing the array spacing decreases the average thresholds below
the threshold values of the arrays in the onsite limit.  
The expected behavior of the average threshold value as compared to the
clean limit, discussed above, is found.
Only at the largest array spacing ($d/r^{isl} = 10$) studied do we  
recover
the dependence of the fluctuations of the threshold voltage
on array length predicted by Middleton and Wingreen\cite{Middleton93}, 
see
inset of Fig.~3(c).

\section{FLOW OF CURRENT}
In this section we discuss the three main voltage
regimes which can be distinguished
in the I-V curve. At voltages very close to the threshold, we show
that the current is linear in $(V-V_T)$. Contrary to the short-range 
interaction case, the slope of this linearity 
depends, not only on the junction resistance and
on $\alpha$, but also on the number of islands and on $d/r^{isl}$. For given
$\alpha$, $N$,  $d/r^{isl}$ and $\{R_i\}$, in the presence of charge 
disorder, the slope of the I-V curve close to threshold
can depend on the potential disorder configuration. 
At intermediate voltages steps in the current are smoothed compared to the 
onsite interaction case. A linear I-V with an offset voltage, closely related 
to the one found for short range interactions, 
characterizes the 
high voltage regime.
\begin{figure}
\includegraphics[width=3.3 in]{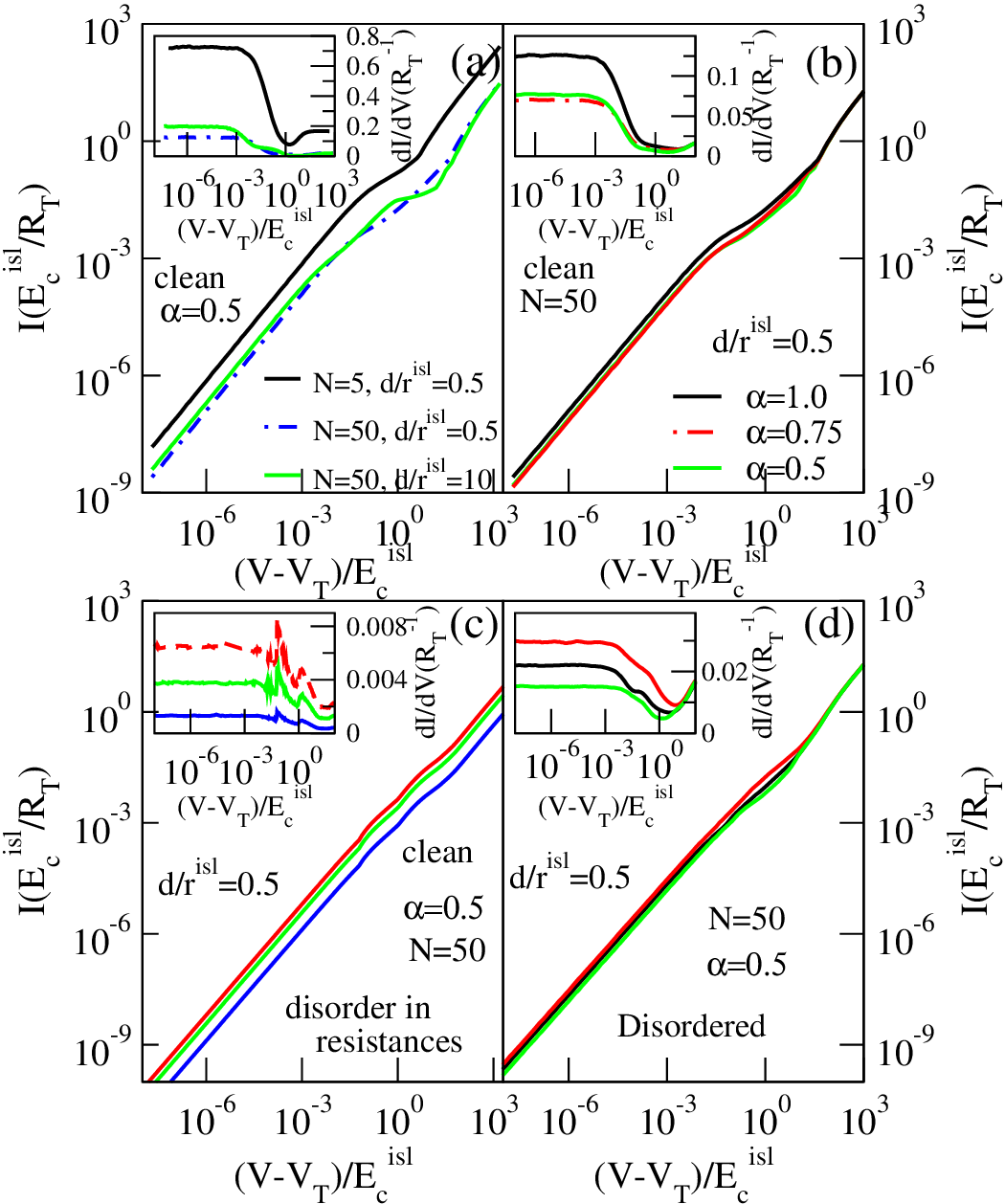}
\caption{
I-V curves of ordered arrays at voltages very close to threshold. The insets 
show the derivatives of the I-V curves. 
Similar to what was found for onsite interactions very close to threshold the 
I-V is linear. The linear regime ends at
voltages $V-V_T\sim 10^{-2} E_c^{isl}$, much lower than the values used in 
experiments to check the power-law dependence close to threshold.
(a) shows how varying the length and 
spacing of symmetrically biased arrays affects the slopes of the linear 
regimes, due to the change in the polarization potential drop factors 
$\Lambda^\alpha_i$ and correspondingly the fraction of potential which drops 
in the
junction which acts as bottle-neck. (b) In a similar way, the slope depends 
on how symmetrically is the array biased.
(c)plots the I-V curves and derivatives corresponding to clean arrays with
equal length, $d/r^{isl}$ and $\alpha$ but different junction resistances.
Resistances in these plots vary randomly in between $(5-11) R_T$, $(8-21) R_T$ 
and $23-83 R_T$ in top, middle and lower curves. 
(d)I-V curves corresponding to three different realizations of 
charge-disordered arrays with all junction 
resistances equal $N=50$ and $\alpha=0.5$. Contrary to what was found for 
onsite interactions the slope of the low-voltage linear current can be 
different for different arrays with the same nominal parameters if there is
charge disorder. This reflects that the bottle-neck is not necessarily a 
contact junction but can be a junction in the bulk.
}
\end{figure}

\subsection{Voltages close to threshold}
For the onsite interaction case we recently resolved the 
controversy\cite{Middleton93,Reichhardt03,Kaplan03,Jha05} on the 
power law dependence of current on $(V-V_T)$, showing that very close to 
threshold it is linear\cite{ourpaper}. Such a linearity is due to the 
bottle-neck character 
of one of the junctions and the linear dependence of the energy for tunneling 
on the bias voltage. These two assumptions remain valid for long-range 
interactions, so the linear $(V-V_T)$ dependence is also found in this case.
In the case of clean arrays, the threshold and low-voltage
bottle-neck for current are found at the contact junctions and 
\begin{equation}
I\sim \frac{\Lambda^{\alpha}_{1,N+1}}{R_{1,N+1}}(V-V_T)
\end{equation} 
Except for $\alpha=1/2$, for which $\Lambda^{1/2}_1/R_1$ and 
$\Lambda^{1/2}_{N+1}/R_{N+1}$ have to be added in the expression for the slope.
 Dependence in $N$, $d/r^{isl}$ and
$\alpha$ via the dependence of 
$\Lambda^{\alpha}_{1,N+1}$ on these parameters is found as seen in Fig.4 (a) 
and (b).
The value of $\Lambda^{\alpha}_i$ which appears in the expression of $V_T$ is 
the same that controls the linearity of the IV curves very close to theshold. 
The 
behavior of the slope of $I$ vs $(V-V_T)$ with $\alpha$, $N$ and $d/r^{isl}$ 
is 
opposite to the one of $V_T$. 
For $\alpha$ different from $1/2$, increasing $N$ and decreasing $d/r^{isl}$ 
decreases the slope because these changes reduce the fraction of the 
polarization potential that drops across the contact junctions. Biasing the
array in a more assymetric way (increasing $(|\alpha-1/2|)$ changes the 
slope 
as the  fraction of the polarization potential that
drops across the contact junction that serves as a bottle-neck at 
small voltages is modified. The slope does also depend on the junction 
resistances, similar to the dependence found for onsite 
interactions\cite{ourpaper}, see Fig.~4(c).

The charge disordered case is to some extent different. The above threshold
bottle-neck (and below threshold current-blocking) junction is not necessarily
either of the contact junctions, so the slope of the linear dependence can be 
controlled by $\Lambda^\alpha_i$ with $i\neq 1,N+1$. The junction that acts as 
the bottle-neck
depends on the particular disorder configuration. This is shown in Fig.~4(d),
which
plots the current and its derivative with respect to voltage for several 
configurations of
disorder corresponding to the same value of $d/r^{isl}$, $N$ and $\alpha$ and
the same junction resistances.
A similar dependence of the slope on the charge disorder configuration was not
present in the short-range case, as  the energy gain for tunneling is 
directly 
modified by changes in the bias potential only at the contact 
junctions, at least for $\alpha=1/2$.    
 The 
slope is generally larger (smaller) when the bottleneck junction lies closer to
the 
edge (middle) of the array. Changing $\alpha$  also modifies the slope of
the disordered case, not shown. 
This modification can be due to the change of $\lambda_i^\alpha$ 
with or without a change in the bottleneck junction.

\subsection{Intermediate voltage regime}
As in the onsite case\cite{ourpaper} the linearity of the current
disappears when the bottle-neck description stops being valid. 
This happens at very small values of $(V-V_T) \sim 10^{-2} E_c^{isl}$.
It is easy to 
show how this situation leads to sublinear behavior. Assume that the transport 
happens through a sequence of $N+1$ tunneling processes and consider a 
bottle-neck process with rate $\Gamma_i=R^{-1}_i \Lambda_i^\alpha \tilde V$ 
with 
$\tilde V=V-V_T$ and another process in the sequence with rate 
$\Gamma_j=R^{-1}_j(E^T_j + \Lambda_j^\alpha \tilde V)$. Here $E^T_j$ is the 
gain in 
energy of the second process at $V=V_T$. If these two processes have rates 
much smaller than the rest of processes in the sequence, the current can be 
approximated by 
\begin{figure}
\includegraphics[width=3.3 in]{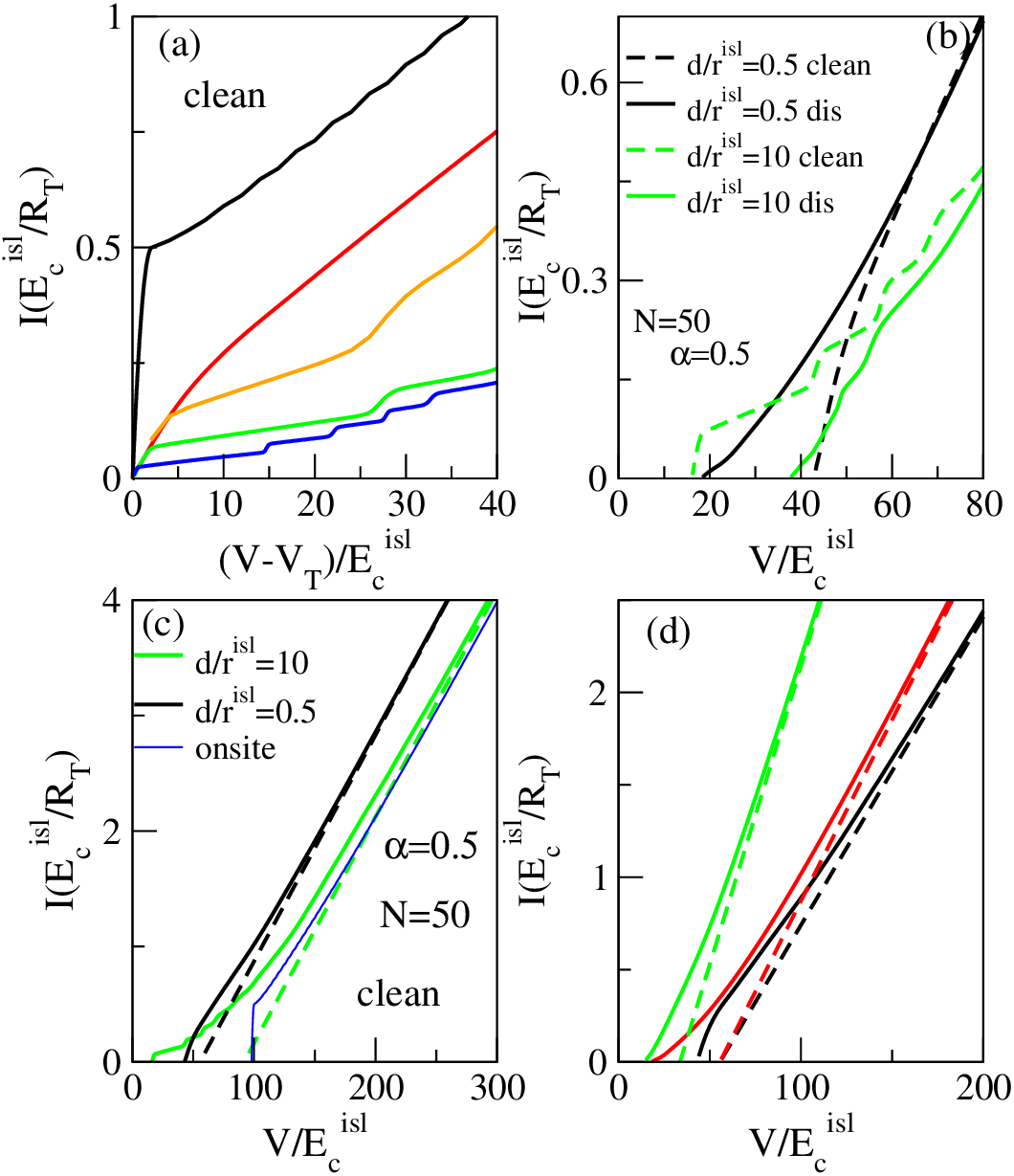}
\caption{In (a) and (b) the I-V characteristics show the 
Coulomb staircase and correspond to clean arrays with 
homogeneous resistances. (a) shows five curves for  different interaction 
ranges and lengths. From top to bottom, the first four 
curves correspond to $\alpha=0.5$ and respectively to  $N=50$ with onsite 
interaction; $N=50$ with $d/r^{isl}=0.5$; $N=30$ with $d/r^{isl}=10$; $N=50$ 
with
$d/r^{isl}=10$. The lowest curve corresponds to a $N=50$ and 
$\alpha=1$ and $d/r^{isl}=10$. The threshold voltage has been substracted. 
$V_T$ equals 98, 43, 14, 16 and 9.7 respectively.  
In the onsite case it is necessary to create a charge gradient 
at every junction to allow current, but once charge can reach the opposite 
electrode it flows easily and the I-V shows a large jump close to threshold. 
For $d/r^{isl}=10$ a charge gradient is not created 
and bulk junctions slow the current flow producing the staircase structure. 
The step width is smaller for forward bias $\alpha=1$ than for symmetric bias, 
but contrary to the behavior found in the Coulomb staircase for onsite 
interactions, with 
long-range interactions the step width is not fixed. The steps are washed out 
for $d/r^{isl}=0.5$ due to a more homogeneous polarization potential drop 
through the array. (b) I-V curves for disordered arrays with 
$d/r^{isl}=0.5,10$. The clean case is included for 
comparison. (c) I-V in a large
voltage scale for clean arrays with homogeneous resistances , from top to 
bottom solid lines correspond to
$d/r^{isl}=0.5$, $d/r^{isl}=10$ and onsite interactions. The dashed lines 
give the asymptotic predictions for $d/r^{isl}=0.5$ and $10$. At high voltages 
all the curves have the same slope given by $R^{-1}_{sum}$. 
The offset voltages differ as the excitonic energies do. In spite of the very 
different threshold voltage and the different dependence  on voltage close to 
threshold, the  $d/r^{isl}=10$ I-V curve 
differs little from the onsite one at high voltages. (d) I-V curves in a 
large voltage 
scale corresponding to $d/r^{isl}=0.5$ symmetrically biased arrays. From top to
 bottom $N=30$ and $N=50$  disordered arrays with homogeneous 
resistances and a $N=50$ clean array with the first resistance ten times 
larger than the other ones. Dashed lines give the asymptotic predictions. The 
slope of both $N=50$ curves differ, but their offset voltages are the same. 
}
\end{figure}

\begin{eqnarray}
\nonumber
I \sim \frac{1}{\tau_i +\tau_j}=\frac{R_i^{-1}\Lambda_i^\alpha\tilde V}{1 + 
\frac{R^{-1}_i\Lambda_i^\alpha \tilde V}{R^{-1}_j(E^T_j+\Lambda_j^\alpha \tilde
 V)}}
\\
\sim R^{-1}_i\Lambda_i^\alpha \tilde V \left ( 1 - \frac{R^{-1}_i 
\Lambda_i^\alpha \tilde V}
{R^{-1}_j E_j^T}\right )
\label{sublinear}
\end{eqnarray}

The slope of the current and the lost of linearity depends on the resistance 
of the junctions, as was also seen in the onsite case\cite{ourpaper}.
In the clean long-range case
comparing the values of $\Lambda^\alpha_i$ 
the linear behavior lasts longer in shorter arrays, smaller 
$d/r^{isl}$ and smaller $\alpha$, as in these cases the values of 
$\Lambda^\alpha_i$ are
more homogenous throughout the array.  
The disordered long-range case is more complex. Due to the non-homogeneous
increase in polarization
voltage drop a junction which has a small energy gain can increase 
this gain more than other junctions
 when the
applied voltage increases and the dependence of the slope with the array 
parameters is not so easily predicted.

If the value of the voltage is increased further several tunneling processes 
are
energetically allowed at each step in a sequence and the discussion of
transport becomes more complex due to the multiple choices available and the
polarization potential drop at the bulk junctions.
Above the linear regime there is a region of smoothed steps in the I-V 
curve. Decreasing $d/r^{isl}$ smooths the steps and for small 
$d/r^{isl}$ they are hardly distinguishable. This behavior is seen in 
Fig.~5(a) 
which compares the onset of current, at voltages not extremely close to 
threshold, for several array parameters. For clarity the curves have been 
plotted as a function of $(V-V_T)$. The staircase profile differs in all these
cases. The top curve corresponds to an $N=50$ array with short-range interactions, 
previously studied\cite{ourpaper}. In this case to allow current flow a 
charge gradient at each bulk junction has to be created, but once charge can
enter the array it flows easily through it. This is reflected in the sharp 
onset of the current close to threshold. The steps at higher voltages are just 
barely visible at this scale. Once the polarization voltage at each junction 
is finite it is not necessary to create a charge gradient at the inner 
junctions and the steps' shape is modified.  The three bottom curves 
correspond to $d/r^{isl}=10$. Several features can be appreciated in these 
curves. The step shape is clearly seen. As the potential drop at the inner 
junction is small, bulk junctions control the flow of currents at each plateau 
and their slope is small. The slope is larger for shorter arrays, and the step 
width depends on the value of $\alpha$. The dependence of the step width on 
$\alpha$ was also found for onsite interactions, but contrary to the short 
range case, for finite $d/r^{isl}$ the step width is not a constant throughout 
the
 curve as the presence of charges in the array influences the energy cost to 
add charges from the electrodes, to the first or last island. As also seen in 
this figure, for small $d/r^{isl}$ the polarization potential drop at the inner
 junction is larger and similar to what happened in the onsite case (but for 
reasons to some extent different) once the charge enters the array it can flow 
easily. In (b) we can see the different I-V curves in clean and disordered 
arrays. Specially interesting is the disordered $d/r^{isl}=0.5$ I-V characteristic. It looks 
superlinear, similar to what would be found if a power-law larger than unity 
is present at these voltages. We have observed that this approximate 
superlinear type dependence is common in disordered arrays with small 
$d/r^{isl}$. If experimentally the power-law behavior expected close to 
threshold is measured at these voltages (larger than those at which the linear
behavior is predicted) the exponent of the power-law could be erroneously 
assigned a value larger than one.

\subsection{Linear behavior at high voltages} 
At very high voltages, in the onsite interaction case we 
showed\cite{ourpaper} that the current can
be approximated by the asymptotic expression
\begin{equation}
I_{asympt} \sim \frac{1}{R_{sum}} (V-\sum_{i=1}^{N+1} E_i^{e-h})
\label{highcurrent}
\end{equation}
with $R_{sum}=\sum_{i=1}^{N+1}R_i$.
The arguments, based on a uniform tunneling rate through all the junctions in 
this voltage regime, 
which led to this expression remain valid in the long-range 
case, with the only change of the quantitative value of the excitonic energies 
$E_i^{e-h}$. The slope of the current does not depend on the range of the
interaction or the presence of charge disorder but the offset voltage at which 
the asymptotic expression cuts the
zero current axis, does\cite{ourpaper,Bakhalov91,Schon}. Confirmation of the 
validity of (\ref{highcurrent}) 
is seen in Figs. 5(c) and (d), where numerical results are compared with the 
asymptotic behavior predicted by (\ref{highcurrent}). In Fig.~5(c) the I-Vs for
symmetrically biased arrays with $N=50$ nanoparticles and different interaction
range are plotted. At high voltages slopes are equal but the offset voltage to
which they extrapolate it is not. Note the difference between the value of 
threshold and the one of the offset. In particular the $d/r^{isl}=10$ curve 
has a smaller threshold and larger offset than the $d/r^{isl}=0.5$. In 
Fig.~5(d) the influence of the number of junctions and their resistance in the 
high-voltage current is
reported and shown to be in good agreement with the approximate prediction.

\section{Voltage drop}
In this section we analyze the effect of long-range interactions in the 
voltage drop through the array. We differentiate the same three regimes as in
previous section. Differences found with respect to the onsite case, 
previously analyzed\cite{ourpaper} are mainly due to the polarization 
potential drop at the bulk junctions, which vanishes in the onsite limit but is
finite when interactions between the charges at the islands and those at the 
electrodes are long-range.   

\begin{figure}
\includegraphics[width=3.3 in]{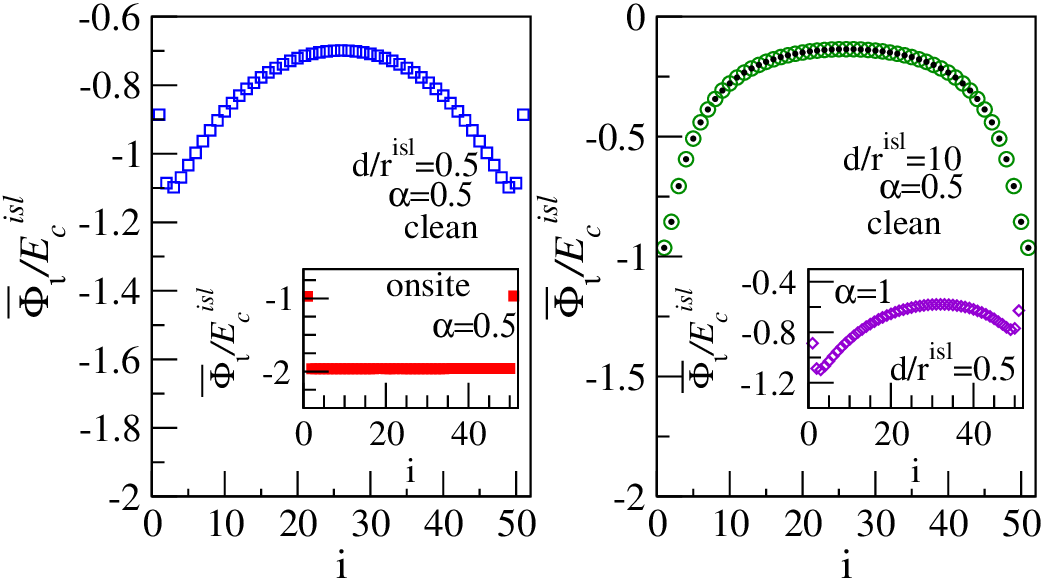}
\caption{Average voltage drop close to threshold for $N=50$ arrays with 
different parameters. Both (a) and (b) correspond to clean arrays. Main 
figures are for $\alpha=0.5$ and long-range interaction. 
The average potential drop essentially equals the polarization potential for 
each value of $d/r^{isl}$, which is plotted as filled small dots in (b) for 
comparison. This behavior contrasts with the potential due to the charge 
gradient which has to be created in the onsite case, shown in the inset of (a).
As shown by comparing main figure in (a) with the inset in (b) a change in the
value of $\alpha$ modifies $\Lambda_i$ and correspondingly the potential drop
through the array.
}
\end{figure}
As discussed in section IV, in a clean array, $V_T$ is given by the 
minimum bias voltage which allows the creation of an electron-hole pair. 
Contrary to the onsite case\cite{ourpaper} there is no charge accumulation in
the array. It is thus expected that very close to $V_T$ 
the voltage drop reflects the polarization drop $\Lambda^{\alpha}_i V$ at each 
junction. The potential drop
distribution thus depend on the length of the array $N$, the bias asymmetry 
$\alpha$ and on the range of the interactions $d/r^{isl}$. This dependence
is confirmed in Figs.~6(a) and (b) 
where the potential drop is plotted for different 
array parameters.  The value of $\Lambda_i^\alpha$ is included for comparison in Fig. 6(b).
The voltage drop is very different from the one found to the onsite case 
(included in the inset of Fig.~6(a)), where in the bulk it is due to charge accumulation 
at the
islands. 
The dependence on the value of the resistance is extremely weak even once the 
polarization potential drop is substracted (not shown) and not observable, 
except if the difference in the value of resistances is very large.
In the disordered case with long-range interaction it is possible that once
charge is allowed to enter the array it can flow. Then the average potential 
drop is approximately the sum of the disorder potential and the polarization 
potential. In general, when this happens the threshold voltage is smaller than 
the one in the clean case as the disorder potential reduces the polarization 
potential drop necessary at at least one of the contact junctions. But for 
large $d/r^{isl}$ is more probable that one or more charges remain stacked in 
the array, similarly to the case of onsite interactions and the charge 
potential due to these stacked charges has to be added.

At intermediate voltages, in the Coulomb staircase regime, we saw in the 
onsite case\cite{ourpaper} that the voltage drop through the array shows an 
oscillatory behavior, with the number of maxima increasing from step to step. 
A similar behavior is found in Fig.~7(a) corresponding to a clean array and 
$d/r^{isl}=10$. In Fig.~7(b) for all the voltages plotted the number of maxima 
is two, and their amplitude decreases until at the largest voltage oscillations
 cannot be discerned. Comparing with Fig.~5 one realizes that the step number 
has not changed. The I-V curves reaches the high-voltage regime without 
showing stepwise behavior. 


At high voltages, the potential drop is qualitatively similar to the one
found in the onsite case\cite{ourpaper}. The voltage drops linearly only after 
subtracting from 
each junction the excitonic energy. The sum of the excitonic energies results
in the offset voltage. The current is equal to $(V-V_{offset})/R_{sum}$ 
and the corresponding voltage drop at each junction is
\begin{equation}
\bar \Phi_i^{high}=R_i I+E^{e-h}_i
\label{drophigh}
\end{equation}  
which satisfies 
$V=\sum_i\bar \Phi_i=I \sum_i R_i+\sum_i E^{e-h}_i=IR_{sum}+V_{offset}$. 
Eq.(\ref{drophigh}) is valid for both ordered and disordered arrays. Some 
examples of 
this behavior are shown in are shown in Fig.~7.  In Fig.~7(c) we can see that 
as
expected, in the absence of resistance disorder, once the excitonic energy is 
substracted the potential drops homogeneously through the array even if there 
is
charge disorder, while it is proportional to the resistance value when 
resistances are not homogeneous, as in Fig.~7(d). 


\begin{figure}
\includegraphics[width=3.3 in]{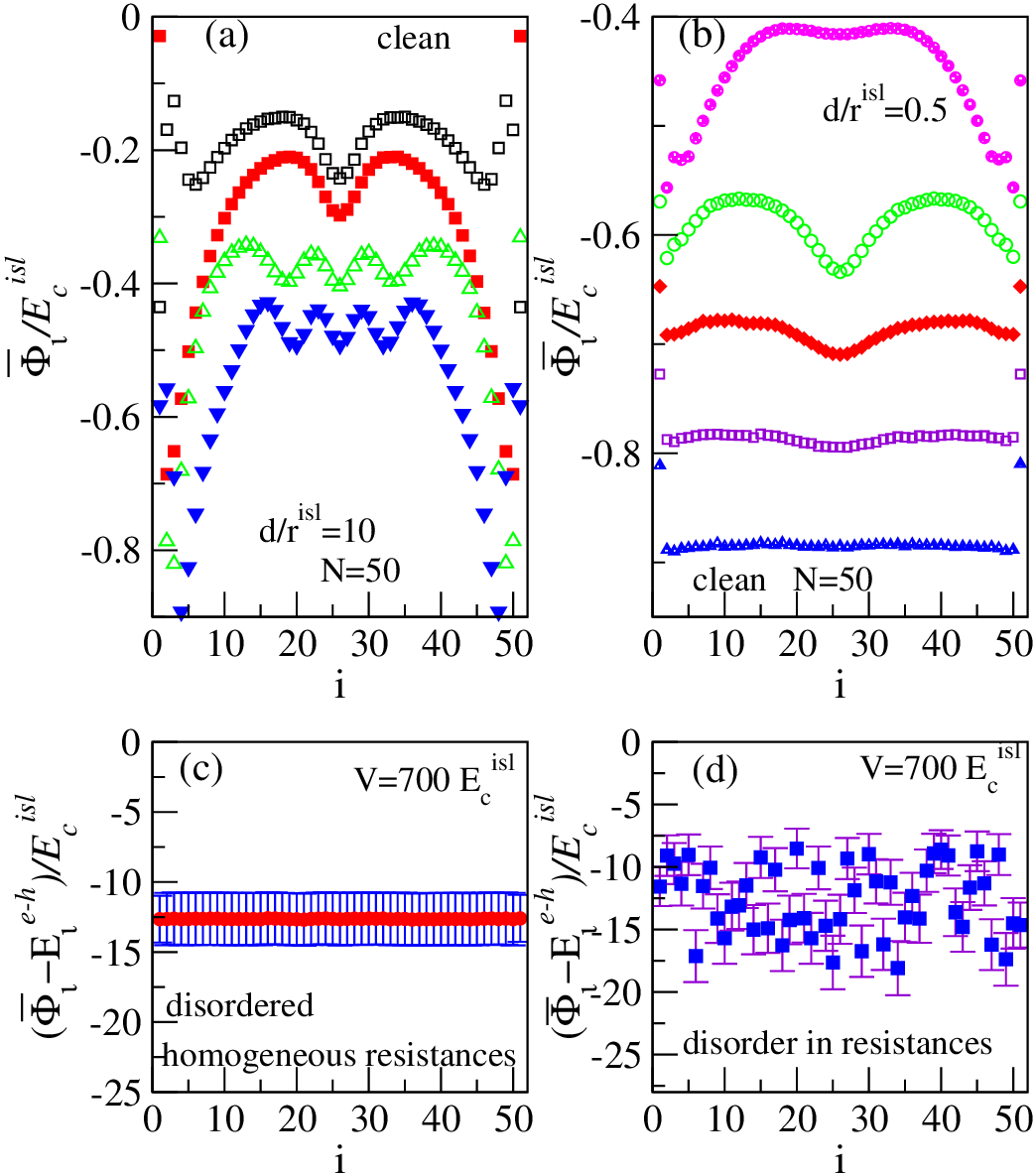}
\caption{Average potential drop at the array junctions at intermediate 
(upper figures) and high 
voltages (lower figures) for $\alpha=0.5$ and $N=50$. Upper figures correspond 
to clean arrays. Curves in (a) are for $d/r^{isl}=10$ and (from top to 
bottom) $V= 20, 32, 44, 56 E_{c}^{isl}$. The oscillations in average potential 
found in the onsite case at intermediate bias voltages are still present, and 
the number of maxima increases in two when going to a higher step. In (b) 
$d/r^{isl}=0.5$ and $V=46,60,70,80,90 E_{c}^{isl}$. The number of maxima has 
not increased in this range of voltages
but oscillations are smoothed with increasing bia voltage. In spite of the 
large 
value of the voltage the current is still in the first step of the Coulomb 
staircase, see Fig. 5(a). The potential 
drop at $V=90 E_c^{isl}$ resembles the one at high voltages, homogeneous except by the 
excitonic term responsible of the offset. In (c) and (d) the excitonic energy 
has been substracted from the average potential drop at high voltages. Once 
this term is substracted the average potential drop is completely homogeneous 
through the array in (c) where there is charge disorder, and all resistances 
are equal but not in (d) which corresponds to a clean array but with 
resistances which vary randomly between $(5-11) R_T$  
}
\end{figure}

\section{Summary}

In this paper we have analyzed the effect of the long-range interaction in the
transport through ordered and disordered nanoparticle arrays. To this end we
have introduced a model which allows a discussion of the relevant quantities
in determining these transport properties. We have introduced the
concept of polarization potential drop, which results from the interaction 
between charges at the electrodes and at the islands. In the model used we take
into account that electrodes are not ideal voltage sources. Their potential
fluctuates in response to tunneling processes (but its effect in the numerical
results is very small). 

We have studied how  the proximity of the nanoparticles modifies the 
$1/r$-interaction due to screening. To determine this screening we have 
developed 
two methods which allow us to calculate the inverse capacitance matrix of the
system under study, see Appendix I. 
As discussed in Appendix II, and shown in Fig. 8(a), 
for the case of metallic nanoparticles, here considered, 
the effect of screening starts to be relevant when $d/r^{isl}\sim 1-2$, there 
is no divergency in the value of $C^{-1}_{ii}$ at small $d/r^{isl}$ values, 
but 
capacitance values and their inverses saturate at finite values. As discussed 
previously by Matsuoka and Likharev\cite{Matsuoka95} 
for cylindrical nanoparticles, the interaction between charges is reduced 
only when the nanoparticles are very close, at larger distances the interaction
increases and approaches the $1/r$ law from above, resulting in a bump in the
interaction potential as compared to  Coulomb law, see Fig. 8(b). 
We have related this anti-screening effect with the dipolar charges induced in
the conductors. 

Long-range interactions screen the disorder potential and induce correlations
between the values of $\phi_i^{dis}$ at different islands. These correlations 
are related to, but smaller than the inverse capacitance matrix elements. The distribution of the island and junction potentials is modified in comparison to the
one found for onsite interactions. These effects are
discussed in section III, and are shown in Figs.~1 and 2.

As discussed in section IV, as in the onsite case, the current is blocked 
up to a threshold value $V_T$.
This threshold value is independent of the resistances of the junctions. 
In clean arrays, $V_T$ is the minimum voltage that allows the creation of an
electron-hole pair, see Figs.~3(a) and (b). 
This behavior is opposite to the one found in the 
onsite case where a charge gradient at the
junctions has to be created to allow the flow of current.       
With long-range interactions, the polarization potential drop across the
array creates a potential gradient which facilitates charge flow. 
In the disordered case, two effects compete that can increase or decrease
the threshold voltage compared to the clean case. Charge accumulation can
be induced by up-steps in the disorder potential, increasing $V_T$, and the
disorder potential distribution can reduce the energy to create an 
electron-hole pair decreasing it. The latest effect dominates for 
small $d/r^{isl}$, see Fig. 3(c).

The current varies linearly with respect voltage close to threshold, in 
the bottle-neck 
regime. This is discussed in section V and plotted in Fig.~4. The slope 
depends,
as the threshold voltage 
does on  the polarization potential
drop, and also on the resistance of the bottle-neck junction. At intermediate
voltages, we find steps in the current, but these are smoothed as compared to 
the onsite case, see Fig.~5(a) and (b). 
Contrary to the onsite case, due to the interaction between the charges in 
different islands the width of the steps is not fixed.
A linear dependence of current on 
voltage  is recovered for large biases, but as in the onsite case, current 
extrapolates to zero at a finite offset voltage (Fig.~5(c) and (d)). The offset
 voltage is given by the sum of the excitonic energies of all the 
junctions, and its value depends on the range of the interaction, but as in 
the case of onsite interactions the slope
of the current is given by the sum of the resistances in series and is
independent of $d/r^{isl}$.

The voltage drop through the array close to threshold, reflects the 
polarization contribution $\Lambda^{\alpha}_{i}$ due to the electrodes, as 
shown in Fig.~6. In the onsite case, the potential drop
in this bottle-neck regime was due to the charge accumulation at the islands, 
necessary to create a potential gradient through the array.

Shown in Fig.~7, at intermediate voltages an oscillatory voltage drop through 
the array, 
similar to the one found for short range interactions is found for large 
values of $d/r^{isl}$. For small $d/r^{isl}$ there is some remnant of this 
behavior but the amplitude of the oscillations can vanish while still being in
the first step of the Coulomb staircase, which is not well defined in this 
case.  
At large voltages, the potential drop is analogous to the one found in 
the onsite case.

In conclusion, we have found that the long-range character of the interaction
modifies the I-V characteristics of one-dimensional arrays for both ordered
and disordered arrays. This modification is greatly due to the finite 
polarization potential drop at the bulk junctions, due to the interactions 
between charges at the islands and at the electrodes. Differences
in transport as a function of the range of the interaction are larger at
low voltages. In both the long-range and onsite cases the current is blocked up
to a threshold voltage $V_T$ and the current depends linearly on for very 
small 
$(V-V_T)$, but the value of both $V_T$ and the slope of the current are
different in both cases, as the mechanism to create a potential gradient 
through the array differs. At high voltages, for given values of the excitonic
energies the transport in both cases is analogous.

\section{APPENDIX I: METHODS TO COMPUTE THE CAPACITANCE MATRIX}

In this appendix we discuss the two methods that we have used to calculate
the interaction strength $C^{-1}_{\alpha \beta}$. In the first one we determine
 the
capacitance matrix using the method of images. This is an iterative method 
which a priori can be used to determine 
the capacitance matrix $C_{\alpha\beta}$ for any geometric configuration of 
spheres, so it is 
valid also for two-dimensional arrays, for example. 
Although 
the algorithm for generating images is straightforward, the number of images 
required to calculate $C_{\alpha \beta}$ makes the numerical 
implementation of this technique nontrivial. While computer memory problems 
can be solved, the computation time is too large to tackle those cases with 
very large arrays and electrodes and small distance between conductors. On the 
other hand, the capacitance matrix is calculated to a given accuracy. Small 
errors in $C_{\alpha\beta}$ can be enhanced and uncontrolled in 
$C^{-1}_{\alpha\beta}$. In the second method the interaction matrix 
$C^{-1}_{\alpha\beta}$ is calculated directly taking into account the symmetry
of the system and the properties of spherical harmonics. It requires the 
inversion of a matrix which can be quite large. It is specially useful and fast
for systems with azimuthal symmetry as the one considered here. Results 
obtained with both methods are in extremely good agreement.

\subsection{Image Charges Method}

The method of images is based on the relation between charges and potentials 
in capacitively coupled conductors. The charge $Q_\alpha$ induced on a 
conductor in 
the presence of $K$ equipotentials at potentials $V_\beta$ is given by the 
capacitance matrix $C_{\alpha\beta}$
\begin{equation}
Q_\alpha=\sum_{\beta=1}^{K}C_{\alpha\beta}V_\beta
\label{cap}
\end{equation} 
The inverse capacitance matrix which enters the free energy (\ref{freecharge}) 
is the inverse of the capacitance matrix $C_{\alpha \beta}$.
We have calculated the capacitance matrix using some properties of spheres and 
the fact that the charge in a conductor $\alpha$ produced by a unit potential 
in a conductor $\beta$ is equal to the capacitance matrix element 
$C_{\alpha \beta}$. We have obtained $C^{-1}_{\alpha \beta}$ inverting 
$C_{\alpha \beta}$.
The method of images\cite{embook} is the placement of imaginary charges inside 
the spheres at positions that make the potential everywhere on the surface of 
the conductor equal a constant. 

To determine the positions of the image charges, we exploit two properties of 
spheres. First, the center of the sphere is equidistant from all points on the 
surface of the sphere. Using this property, the surface of an sphere of radius 
$R$ can be set to a potential $V$ by placing an image at the center of the 
sphere of charge $q=VR$. Second, for every point outside a sphere there is a 
point inside the sphere for which the ratio of the distances between these two 
points and any point on the surface of the sphere is a constant. From here it 
follows that if a charge $q_R$ is located at the outside point, at a distance 
$d_c$ from the center, an image charge $q_I$ placed at the inside a distance 
$R^2/d_c$ 
from the center in the radial 
line, with charge
\begin{equation}
q_I=-q_R\frac{R}{d_c}
\end{equation}
will set the potential to zero everywhere on the surface of the sphere.
We determine the $(N+2)\times(N+2)$ capacitance matrix, column by column, by 
determining the set of image charges that sets the potentials of the spheres 
to $V_\alpha=\delta_{\alpha\beta}$. The capacitance matrix elements 
$C_{\alpha\beta}$ are given by the sum of all the charges in sphere $\alpha$. 
To 
set the potential of the $\beta$ sphere, with radius $R_\beta$ to one, we 
place a charge with magnitude $R_\beta$ at the center of this sphere 
$x_{\beta}$. The remaining spheres are grounded by placing images inside each 
sphere with charges 
\begin{equation}
q_\nu=-\frac{R_{\nu}q^{old}}{|x_{q_{old}}-x_{\nu}|}
\end{equation}
at positions
\begin{equation}
x_{q_\nu}=x_{\nu}+\frac{R^2_\nu}{x_{q_old}-x_\nu} \; \mbox{.}
\end{equation}
Here $q^{old}$ and $x_{q_{old}}$ are the value and the position of the charge 
which creates the inhomogeneous potential that we want to compensate and 
$R_\nu$ and $x_\nu$ are the radius and position of the center of the sphere to 
which we add the image charge $q_\nu$. These image charges are added to all 
the spheres except the one in which $q_{old}$ is placed. The charges that have 
been added generate new inhomogeneous potentials at the rest of the spheres 
and have to be compensated following the same method. This process repeats 
iteratively for all the charges added to all the spheres.
During each 
iteration $n$, the number of new images required to compensate the potential 
of the other spheres approximately equals $(N+1)^n$. We eliminate some of the 
images by discarding images with a magnitude that is smaller than a suitable 
cutoff value, $q_{cutoff}$. We required $q_{cutoff}$ to be small enough that 
the relative diffferences between the matrix elements generated with the 
cutoff value $q_{cutoff}$ and by a larger cutoff value 
$q'_{cutoff}=10 q_{cutoff}$ are less than one percent.

\subsection{High-order multipoles method} 
Following Wehrli {\it et al}\cite{Wehrli},  the energy of the system, given by 
(\ref{freecharge}) can be rewritten in terms of the higher-order multipolar 
charges induced by the charges on the conductors as 
\begin{equation}
F=\frac{1}{2} \sum_{\alpha, \beta, l, m, l', m'}Q^{\alpha,*}_{l, m}G^{\alpha
  \beta}_{l,m,l',m'}Q^{\beta}_{l' m'} \; \mbox{.}
\label{freemultipolar}
\end{equation}
Here Greek indices denote the conductors, $l$ and $l'$ denote the order of the 
multipole,
and $m= -l,\ldots, l$ and $m'=-l', \ldots, l$ denote the azimuthal number. 
This matrix $G$ is hermitian with
respect to the exchange of $\alpha,l,m$ and $\beta,l',m'$. Using the linear 
response form for the induced multipoles, the higher-order multipolar charges, 
$Q^\alpha_{l,m}$, can be expressed in terms of the (monopolar) charges on the 
conductors $Q_\gamma=Q^{\gamma}_{00}$, as
\begin{equation}
Q^\alpha_{l,m}=\sum_{\gamma} \Gamma^{\alpha \gamma}_{l,m}Q_\gamma \; \mbox{.}
\label{qlm}
\end{equation}
Substituting (\ref{qlm}) into (\ref{freemultipolar}) and comparing it with 
(\ref{freecharge}), the inverse capacitance matrix can be expressed as
\begin{equation}
C^{-1}_{\gamma \eta}=  \sum_{l,m,l',m',\alpha,\beta}G^{\alpha
  \beta}_{l,m,l,m'}\Gamma^{\alpha \gamma *}_{l,m}\Gamma^{\beta\eta}_{l',m'} \; \mbox{.}
\end{equation}
The multipolar charge induced is the one which minimizes the energy. 
Separating the 
monopolar contribution ($l,m=0$) in the expression of the free energy, and 
minimizing 
the latter with respect to $Q_{l,m}^\alpha$, we obtain
\begin{equation}
{\bf Q_A} =-\hat G^{-1}_{AB}\hat G_{B0}{\bf Q_0} \; \mbox{.}
\end{equation}
Here $A=l,m$ and $l \neq 0$, correspondingly $B$, and the equation is written 
in vectorial and matrix notation. 
Once this expression is substituted in the free energy, the  inverse 
capacitance  can be written in
terms of the $\hat G$ matrices
\begin{equation}
\hat C^{-1}=\hat G_{00}-\hat G_{0A}\hat G^{-1}_{AB}\hat G_{B0}
\label{correction}
\end{equation}
The order of approximation in this method is the number of the highest 
multipoles $l,l^\prime$ included.
Matrix $\hat G_{00}$ has dimension $N_s \times N_s$ with $N_s$ the total 
number of conductors. Matrices
$\hat G_{0A}$ and $\hat G_{B0}$ are $N_s \times (N_s N_{totalmulti})$ and 
($N_sN_{totalmulti}$)$\times N_s$
respectively, and matrix $\hat G_{AB}$ has dimension 
$(N_sN_{totalmulti})\times (N_sN_{totalmulti})$. 
$N_{totalmulti}$ is the maximum number of multipolar terms considered. 
Formally it is
\begin{equation}
N_{totalmulti}=\sum_{l=1,l_{max}}(2l+1)
\end{equation}
with $l_{max}$ the order of the maximum multipole included in the
approximation. However the symmetries of the problem can help us to reduce it
as the $\hat G_{AB}$ elements corresponding to certain $m_l$ can be seen to
vanish by symmetry. In particular in the case of azimuthal symmetry, 
considered 
in the text, only
$m=0$ gives non-zero values and the number of terms included can be reduced to
$N_{totalmulti}=l_{max}$. Depending on the geometry of the conductors it can 
be 
convenient to use different number of $l_{max}$ for different conductors. In 
particular
in the case of an array of small islands sandwiched by two large electrodes, 
it is
better to use a larger number of multipoles at the electrodes. Most of the 
cases presented here are done with $l_{max} \sim 8$.

The expression for $G^{\alpha \beta}_{lm}$ follows from the decomposition of 
$1/|{\bf a} -{\bf b} -{\bf R}|$, with ${\bf a}$, ${\bf b}$ and ${\bf R}$ three
points in space and depends on the geometry of the conductors.
 For $\alpha\neq \beta$

\begin{eqnarray}
\nonumber
G^{\alpha \beta}_{l_1 m_1 l_2 m_2}=\left[ \frac{(l_1 + l_2
    - m_1 - m_2)! (l_1 + l_2 - m_1 + m_2)!}{(l_1 + m_1)!(l_1
    -m_1)!(l_2+m_2)!(l_2-m_2)!}\right ]^{1/2}
\\
(-1)^{l_2+m_2}I_{l_1+l_2+m_1-m_2}(x_\beta-x_\alpha)
\end{eqnarray}
with $I_{l,m}$ the irregular solid spherical harmonics
\begin{equation}
I_{lm}({\bf r}) = \frac{1}{r^{l+1}}\sqrt{\frac{4 \pi}{2l +1}}Y_{lm}(\Omega)
\end{equation}
The sign of $G^{\alpha \beta}_{l_1 m_1 l_2 m_2}$ depends not only on
$l_2$ and $m_2$, but also on the order $\alpha \beta$ or $\beta \alpha$
through the dependence of $I_{l_1+l_2+m_1-m_2}(x_\beta-x_\alpha)$.

For the case of an sphere $\alpha$ with radius $R_\alpha$,  
$G^{\alpha \alpha}_{l_1,m_1, l_2 m_2}$
\begin{equation}
G^{\alpha \alpha}_{l_1 m_1 l_2 m_2}=\delta_{m_1 m_2}\delta_{l_1 l_2}
\frac{1}{R_{\alpha}}^{2l_1 +1}
\end{equation}
The case of spheres on a row is especially simple. In this case we have
azimuthal symmetry what means that all terms involving $m \neq 0$ should
vanish. Thus at order $l_{max}$ we have just $N_{totalmulti}=l_{max}$.
This simplification allows us to go to reasonably high orders.
 We can
eliminate the indexes $m_1,m_2$ from the matrix $G$. Together with the
diagonal terms $G^{\alpha \alpha}$ calculated above, and using that 
\begin{equation}
Y_{l0}=\sqrt{\frac{4 \pi}{2 l +1}} P_l(cos \theta)
\end{equation}
and $P_l(1)=1$ and $P_l(-1)=(-1)^l$ the equations are greatly simplified.
Thus
\begin{eqnarray}
G^{\alpha \beta}_{l_1,l_2}=\frac{(l_1 + l_2)!}{l_1 ! l_2!} (-1)^{l_1}
\frac{1}{r_{\alpha \beta}^{l_1 +l_2 +1}}, if x_\beta > x_\alpha \nonumber \\
G^{\alpha \beta}_{l_1,l_2}=\frac{(l_1 + l_2)!}{l_1 ! l_2!} (-1)^{l_2}
\frac{1}{r_{\alpha \beta}^{l_1 +l_2 +1}}, if x_\beta < x_\alpha
\end{eqnarray}
for $\alpha \neq \beta$. Here $r_{\alpha \beta}$ is the distance between the
centers of the spheres $\alpha$ and $\beta$. The diagonal of $G_{0A}$ and 
$G_{A0}$
are zero and $G^{\alpha \beta}_{0 A}=G^{\beta \alpha}_{A 0}$. 
Note that 
\begin{equation}
G^{\alpha \alpha}_{00} = \frac{1}{R_\alpha}
\end{equation}
and
\begin{equation}
G^{\alpha \beta}_{00}=\frac{1}{r_{\alpha \beta}}
\end{equation}
The zero-order approximation recovers our expectation for the case of far-apart
spheres. 
 The correction to the inverse capacitance 
due to the higher order
multipoles is given by $-\hat G_{0A}\hat G^{-1}_{AB}G_{B0}$.
As spheres come closer, higher order terms become more and more
important. This is reasonable taking into account that the interaction between
two multipolar charges $Q^\alpha_{l_1 m_1}$ and $Q^\beta_{l_2 m_2}$ decays as 
$r_{\alpha \beta}^{l_1 +l_2 +1}$.

\section{APPENDIX II. Interaction between charges and capacitance matrices}

When two conductors become closer together the mobile 
charges in 
their surfaces  screen the interaction between the charges in them, 
compared to the $1/r$ Coulomb law which describes the interaction between 
two isolated point-like charges. 
Other metallic systems in the
surrounding environment  contribute to this screening. 
It is expected that the charging energy of an 
sphere, or the energy to create an electron-hole pair between two islands will
depend on the distance between the particles in the array. 
In this appendix we describe the modification of the interaction due to 
screening and compare it with other models and calculations available in
the literature.
$C^{-1}_{\alpha \beta}$ is 
calculated as described in Appendix I. Both methods described give the same
value of $C^{-1}_{\alpha\beta}$ to several  digits when the accuracy
 used in the computation is 
good enough. Here, inverse
capacitances are measured in units of $C^{-1}_{isl}$, the value of the inverse
capacitance of a nanoparticle $C^{-1}_{ii}$ when it is isolated. We find that
the effect of screening is essentially restricted to the interactions between 
charges which are in the same islands or in the closest neighboring islands and
only when the particles are very close $d/r^{isl}\leq 1-2$.

\begin{figure}
\includegraphics[width=3.3 in]{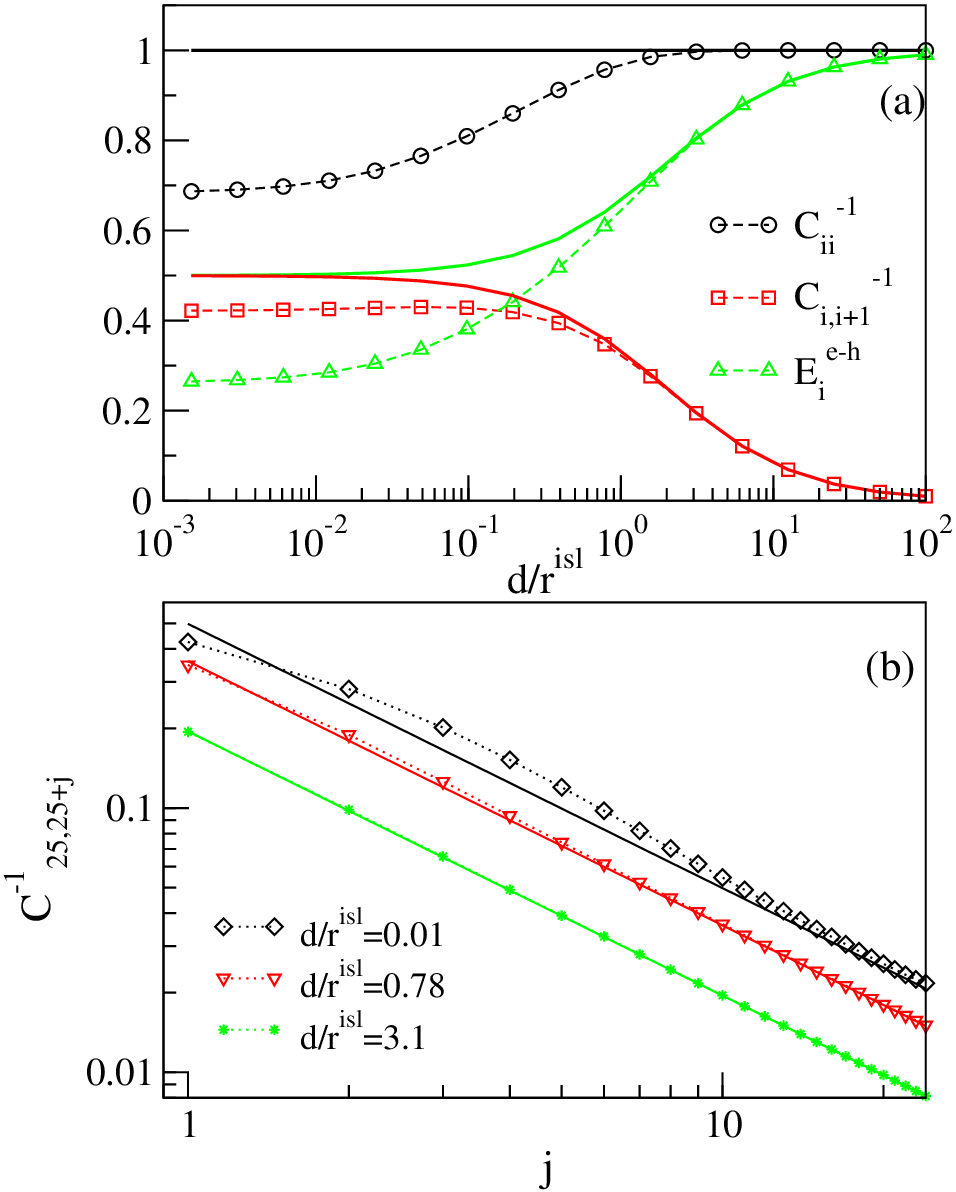}
\caption{(a) Island inverse capacitance $C^{-1}_{ii}$, nearest-neighbor 
interaction $C^{-1}_{i, i+1}$ and excitonic energy, all in units of $(C^{isl})^{-1}$, at the center of a 50 
nanoparticle array (without electrodes) as a function of the interisland  
separation. Solid-lines give the value obtained with a $1/r$ interaction 
between charges. The effect of screening is evident for 
$d/r^{isl}\leq 1-2$. All the plotted quantities saturate to a finite value as 
$d/r^{isl}$ vanishes.(b) Decay of the interaction potential from the center
of a 50 nanoparticle array (without electrodes) as a function of 
island position for different values of $d/r^{isl}$. 
Inverse capacitances are given in units of $(C^{isl})^{-1}$
Solid lines correspond 
to an 
unscreened $1/r$ law. The bump is clearly observed only for very small 
$d/r^{isl}$. It is much weaker than the one found in\cite{Matsuoka95} and is restricted to the first nearest neighbor. For 
large distances the $1/r$ law is
approached from above. The effect of screening is negligible for $d/r \sim 3$.
In (a) and (b) dashed and dotted lines are included as a guide to the eye.}
\end{figure}

Three important quantities for the transport are the island inverse capacitance
$C^{-1}_{ii}$, the nearest-neighbor interaction $C^{-1}_{i, i+1}$ and the
resulting excitonic energy $E^{e-h}_i$.  
In Fig.~8(a) we show how these quantities
depend on the distance between 
the particles, $d/r^{isl}$ for the case of an array with $N=100$ equal-sized 
islands, at the center of the array. 
The effects of screening start to be relevant for $d/r^{isl} < 1-2$. At this
value $C^{-1}_{ii}$ starts decreasing, and when $d/r^{isl} \rightarrow 0$, 
approximately at $d/r^{isl}=0.002$ it saturates at about 0.68 of its bare 
value.
$C^{-1}_{i,i+1}$ increases following a $1/r$ law as $d/r$ is reduced, until 
$d/r^{isl}\sim 1-2$. Then it decreases slightly due to the screening of the 
interaction, and finally it saturates at about $d/r^{isl}=0.1$. At large $d/r^{isl}$ 
the 
$E^{e-h}_i$ starts decreasing when $d/r^{isl}$ is reduced, according to the 
$1/r$ increase of $C^{-1}_{i,i+1}$. For $d/r^{isl}<2$, its value is affected 
both by the increase of $C^{-1}_{i,i+1}$ and the decrease of $C^{-1}_{ii}$, at 
small distance is controlled by this last effect. Finally it saturates.

The modification of the distance dependent interaction in one-dimensional 
arrays
has been previously
studied for some particular island geometries. 
For the case of thin circular disks (with disk axis perpendicular 
to the array axis), 
Whan et al.\cite{Whan96} 
found that a $1/r$ law describes well
the dependence of the non-diagonal inverse capacitance matrix
elements on distance  and that the diagonal elements are reduced.
Likharev and Matsuoka\cite{Matsuoka95} analyzed
 the cases of an array of cubic islands (with
inverse capacitance matrix calculated using the finite-differences software 
FASTCAP) and a continuum model in which the discrete periodic structure is 
replaced by a 
continuous dielectric medium. They found that the 
dependence of 
the interaction on the distance between the charges $r$ has a crossover 
between a linear 
dependence at short distances and a $1/r$ law at long distances. There is a 
bump on the interaction when compared to a $1/r$ law. The 
potential 
approaches the $1/r$ law from above. 
The interaction potential could by approximated by the 
expression
\begin{equation}
\label{fit}
U(m)=\frac{e^2}{a} \left ( \frac{\alpha}{m_0}exp\left ( \frac{-\kappa
    m}{m_0}\right ) + \frac{1}{m}\left [ 1 - exp \left (\frac{- \kappa
    m}{m_0}\right ) \right ] \right )
\end{equation} 
Here  $m$ is the distance in units of the array period, $a$, $m_0=r_0/a$, with 
$r_0=S\epsilon/\pi$ a characteristic decay length of the interaction. 
$\epsilon$, the inter-islands dielectric constant, $S$ the junction 
surface and $\kappa$  a fitting parameter with value very close to 1 and 
related to $\alpha$ by
\begin{equation} 
\alpha=\frac{2}{\kappa}-\frac{\kappa}{2}
\end{equation}

The dependence of $C^{-1}_{i,i+j}$ on $j$ that we have obtained 
for the experimentally relevant case of 
an array of spherical particles 
is plotted in Fig.~8(b). 
Screening is less
important for the case of spheres than for cubic islands but larger than for 
the thin disks analyzed by 
Whan {\it et al}\cite{Whan96}. 
Compared with a $1/r$ law, at short distance the screened 
interaction 
potential decreases and at large distances it increases. When the particles 
are very close, there is a bump in the
renormalization of the interaction. Only when two charges are in the same 
particle or at the nearest neighbor one the interaction between them decreases.
In other cases the interaction increase. 
We have not been able to fit our results with (\ref{fit}). 
We recover the bump found by Likharev and Matsuoka\cite{Matsuoka95}, 
but it is smaller than the one 
they observe, even when our particles are very close. We believe that the 
differences are due to the different geometry of the system under study.

From the decomposition of the induced screening charge in high-order 
multipoles discussed in previous appendix it is possible to get some insight 
on how does the bump appear. 
The correction to the inverse capacitance $\delta C^{-1}$ due to screening 
is $-\hat G_{0A}\hat G^{-1}_{AB}\hat G^{-1}_{B0}$, see (\ref{correction}), 
which 
has the same
sign as the monopole-induced multipole interaction. Thus, 
$\delta C^{-1}_{\alpha \beta}$  has the same sign as the interaction of the 
monopolar charge in $\alpha$ with the multipolar charge induced in all 
conductors by the 
monopolar charge in $\beta$. Let us restrict to dipolar order, 
which will be the largest contribution. $Q_\beta$ generates dipolar charges in 
all the other conductors $\gamma$, equal to 
$Q^{\gamma \beta}=\sum_{\nu}(G^{\gamma\nu}_{11})^{-1}G^{\nu\beta}_{10}Q_{\beta}$
 and the monopolar charge in $\alpha$ interact with these induced charges via 
$G^{\alpha \gamma}_{01}$. Both the monopolar-dipolar interaction $G_{01}$ and 
the dipolar induced charges are odd quantities with respect to position. To 
dipolar order, all the elements of $G^{-1}_{11}$ are positive. The sign of the 
two odd quantities will control the sign of $\delta C^{-1}_{\alpha \beta}$.

Consider  $\delta C^{-1}_{\alpha \beta}$, with
$0<\alpha<\beta<N$, the charges induced by a monopolar charge placed in 
$\beta$ in 
conductors from
$1$ to $\alpha-1$ and from $\alpha+1$ to $\beta-1$ have the same sign, 
but the interaction of a monopolar charge in $\alpha$ with them, have
opposite. Thus those terms coming from the charge induced in conductors from
$1$ to $\alpha-1$ and those from conductors $\alpha +1$ to $\beta -1$ 
contribute to $\delta C^{-1}_{\alpha}$ with different sign. The sign of the 
contribution
of the conductors at the right of $\beta$ will be the same as those at the
left of $\alpha$ and both induced charge and interaction have opposite sign. 
As further are these conductors to $\alpha$ and $\beta$ the correction will be
smaller. Individual contributions from each conductor $\gamma$ will be larger 
when $\alpha < \gamma <\beta$. The contribution of the interaction term which 
comes from the dipoles generated in conductors between $\alpha$ and $\beta$ 
increase $C^{-1}_{\alpha \beta}$.  Only when the contribution of the terms 
which decrease the inverse capacitance matrix element is able to compensate 
the contribution of those ones which increase it, the total change will be 
negative. As closer are $\alpha$ and $\beta$ the number of terms increasing the
 interaction will
decrease. $C^{-1}_{\alpha \beta}$ is expected to be smaller than the bare 
value, only if $\alpha$ and $\beta$ are very close, what results in the 
appearance of the bump and the anti-screening effect at intermediate and large 
distances.

The interaction obtained differs considerably from the one resulting from 
capacitive coupling only to nearest-neighbors. This form of the interaction 
has been used frequently in the  literature, and applies when the system is 
coupled to
a gate electrode which screens the long-range part. 
The capacitance matrix 
has a triband form. Only $C_{ii}=C_0+2C$ and $C_{i,i\pm 1}=-C$ are finite. 
The inverse of $C_{ij}$ for the case of an infinite 
array becomes
\begin{equation}
C^{-1}_{ij}=\frac{1}{C^{\infty}}e^{-|i - j|/\xi}
\end{equation}
with $\xi$ and $C^\infty$ defined from 
$C^{\infty}=2C sh (\xi^{-1}) =\left (C_0^2 + 4CC_0\right )^{1/2}$. Due to the 
exponential dependence $\xi$ can be viewed as the decay 
length of the interaction, which increases with 
$C/C_0$. Interactions on the same island are given by 
$C^{-1}_{ii}=1/C^{\infty}$. In the case of a finite array of length $N$ 
this value is 
approached, from below, as $N$ increases. The onsite case is recovered when 
$C=0$ and long-range interactions appear in the opposite limit 
$C_0/C\rightarrow 0$. In the later limit the interaction potential goes like
\begin{equation}
C^{-1}_{ij}=\frac{1}{2C}\left ( \frac{N}{2}-|i -j|\right) 
\end{equation}
It decays linearly with 
distance. The 
energy to 
create an electron-hole pair 
$E_i^{e-h}=1/2C^{-1}_{ii}+1/2C^{-1}_{i\pm 1,i\pm1}-C^{-1}_{i,i\pm1}$ remains 
bound and equal to $1/(2C)$. On the contrary,
the diagonal element $C^{-1}_{ii}=\frac{1}{4C}N$ diverges with 
the array size. 
There is not such unphysical divergence in $C^{-1}_{ii}$ with the 
array size in our 
model.

To analyze the transport properties the array is sandwiched between two 
electrodes, much larger than each of the nanoparticles. To this end we
consider a one-dimensional array of $N$ nanoparticles placed in between
two large spheres, with radius $R$, which play the role of the leads. The 
large size ensures large screening and
that $C_{00}^{-1}$ and $C_{N+1,N+1}^{-1}$ are much smaller than the islands
$C^{-1}_{ii}$. 
The spherical shape greatly simplifies the calculations of the inverse 
capacitance matrix.
The interaction between the charges at the islands
and those at the electrodes and the inverse capacitance elements of the 
electrodes determine $\lambda_i^\alpha$ and $\Lambda_i^{\alpha}$, which 
control the polarization voltage drop through the array and to a large
extent the current flow at small voltages, see Figs.~6(a) and (b).

For the size of the electrodes used in the text, $R \sim 50-100 r^{isl}$, the
inverse capacitance of the islands close to the electrodes is slightly reduced
compared to those at the center, except for very small $d/r^{isl}$. 
For small $d/r^{isl}$ the inverse capacitance 
of islands at the center of the array is almost insensitive to the presence of
the electrodes.  If the electrodes are much larger the dependence of the
inverse capacitance matrix elements with the size of the electrodes can become
non-monotonous. This behavior, like the one found for small $d/r^{isl}$ is 
most probably associated to the 
spherical shape chosen to model the electrodes.

We thank S. Wehrli for useful discussions.
Financial support from the Swiss National Foundation, NCCR MaNEP of the Swiss 
National Fonds, the Spanish Science and Education Ministry through Ram\'on y 
Cajal contract FPI fellowship and grant No. FIS2005-05478-C02-02 and the 
Direcci\'on General de Universidades e Investigaci\'on 
de la Consejer\'{\i}a de Educaci\'on de la Comunidad de Madrid and CSIC 
through Grant No. 200550M136  is gratefully acknowledged.
Work at UT Austin was supported by the Welch Foundation, by the NSF under 
grant DMR-0606489 and by the ARO under grant W911NF-07-1-0439.

\end{document}